\documentclass[
reprint,
superscriptaddress,
nofootinbib,
 amsmath,amssymb,
 aps, prc,
]{revtex4-2}

\usepackage{graphicx}% Include figure files
\usepackage{dcolumn}% Align table columns on decimal point
\usepackage{bm}% bold math
\usepackage[colorlinks,citecolor=blue]{hyperref}% add hypertext capabilities
\usepackage{orcidlink}
\usepackage{booktabs}
\usepackage{braket}

\newcommand{\dint}{-2.5pt}

\newcommand{\kett}[1]{\ensuremath{|{#1}\rangle\rule{\dint}{0pt}\rangle}}
\newcommand{\bbrakett}[1]{\ensuremath{\langle\rule{\dint}{0pt}\langle{#1}\rangle\rule{\dint}{0pt}\rangle}}
\newcommand{\cev}[1]{\reflectbox{\ensuremath{\vec{\reflectbox{\ensuremath{#1}}}}}}

\begin{document}

\title{Reentrance of proton-neutron pairing in hot nuclear systems}

\author{T. Vu Dong \orcidlink{0009-0005-6916-1253}}
\affiliation{Institute of Fundamental and Applied Sciences, Duy Tan University, Ho Chi Minh City 70000, Vietnam}
\affiliation{Faculty of Natural Sciences, Duy Tan University, Da Nang City 55000, Vietnam}
\affiliation{Bogoliubov Laboratory of Theoretical Physics, Joint Institute for Nuclear Research, 141980 Dubna, Russia}

\author{Alan A. Dzhioev\orcidlink{0000-0003-0642-1079}}
\email{\textcolor{black}{Corresponding author:} dzhioev@theor.jinr.ru}
\affiliation{Bogoliubov Laboratory of Theoretical Physics, Joint Institute for Nuclear Research, 141980 Dubna, Russia}

\author{A. I. Vdovin\orcidlink{0000-0003-2694-9460}}
\affiliation{Bogoliubov Laboratory of Theoretical Physics, Joint Institute for Nuclear Research, 141980 Dubna,  Russia}

\author{N. Quang Hung\orcidlink{0000-0002-9739-301X}}
\email{\textcolor{black}{Corresponding author:} nguyenquanghung5@duytan.edu.vn}
\affiliation{Institute of Fundamental and Applied Sciences, Duy Tan University, Ho Chi Minh City 70000, Vietnam}
\affiliation{Faculty of Natural Sciences, Duy Tan University, Da Nang City 55000, Vietnam}

\date{\today}

\begin{abstract}

We develop a generalized finite-temperature proton-neutron BCS (FT-pnBCS) framework using the superoperator formalism, incorporating both isovector  and isoscalar monopole pairing channels. Numerical calculations for a schematic equidistant multilevel model and realistic even-even Ge isotopes demonstrate the emergence of proton-neutron ($pn$) pairing reentrance in even-even asymmetric  ($N>Z$) nuclei with preexisting like-nucleon pairing  correlations. This non-monotonic behavior arises from thermal excitations that partially lift Pauli blocking of single-particle orbitals near the chemical potentials, thereby enlarging the phase space for $pn$ pair formation.  We uncover a delicate interplay between thermal unblocking and like-nucleon pairing, which can either suppress or enhance $pn$ correlations depending on temperature and shell filling.  A qualitative analysis of Fermi charge-exchange strength functions in hot $^{72}$Ge, which neglects the residual interaction between thermal quasiparticles,  suggests that $pn$ pairing reentrance may alter the transition strength distribution around $T\approx1$~MeV. This indicates that finite-temperature $pn$ correlations could  potentially impact  stellar weak-interaction rates in  $rp$-process and supernova environments.
\end{abstract}

\maketitle

\section{INTRODUCTION}

 The pairing interaction lies at the heart of nuclear structure physics~\cite{Broglia2013}.  While like-nucleon pairing between protons (\(pp\)) or neutrons (\(nn\)) dominates in neutron-rich nuclei (\(N \gg Z\)), proton–neutron (\(pn\)) pairing becomes significant in nuclei with nearly equal numbers of protons and neutrons (\(N \approx Z\))~\cite{PPNP78_Frauendorf}, where valence protons and neutrons occupy closely spaced orbitals. Such \(N \approx Z\) nuclei are typically located near the proton drip line and lie along the astrophysical rapid proton-capture (\(rp\)) process path. Consequently, the study of \(pn\) pairing is not only essential for a complete understanding of nuclear structure but is also expected to play a crucial role in nuclear astrophysics~\cite{kaneko2005proton}.

At zero temperature, the simultaneous treatment of like-nucleon and $pn$ pairing correlations has been extensively investigated, primarily within the $pn$ Bardeen-Cooper-Schrieffer (pnBCS) framework. The isovector ($t=1$) pnBCS formalism was first derived by Goswami in 1964~\cite{goswami1964treatment}. This approach was subsequently extended to incorporate both isovector and isoscalar ($t=0$) correlations between nucleons in time-reversed states~\cite{chen1967generalized,goodman1968restoration}. A fully isospin-generalized theory, which additionally accounts for pairing correlations in identical orbitals, was later developed by Goodman~\cite{NPA186_Goodman}.

In hot nuclei, the behavior of $pn$  pairing has been explored in only a limited number of studies. In Ref.~\cite{kaneko2005proton}, \(pn\) pairing energies in odd–odd \(N = Z\) nuclei were extracted from a “thermal” odd–even mass difference based on shell-model calculations. These results indicate an enhancement of \(pn\) pairing correlations at finite temperature in such nuclei. Similarly, Ref.~\cite{sheikh2005reappearance} investigated both  \(t = 1\) and \(t = 0\) pairing correlations in rotating  deformed  $N=Z$  systems at finite temperature using an exactly solvable Hamiltonian. Their findings reveal that the \(pn\) pairing gap can increase when the nucleus is simultaneously hot and rotating. In both cases, the enhancement of $pn$ pairing at finite temperature can be understood as a consequence of the thermal unblocking of single-particle levels near the Fermi surface, which would otherwise remain blocked at low temperatures either by an unpaired nucleon or by nucleons from pairs broken by rotation. Thermal excitations promote nucleons to higher levels and increase the number of proton-neutron configurations available for pair formation. As a result, pairing correlations that are suppressed at zero temperature may become enhanced over a certain temperature interval.

In asymmetric nuclear systems, $pn$ pairing correlations are strongly suppressed in the ground state by Pauli blocking, as excess neutrons occupy single-particle orbitals that would otherwise be available for $pn$ pair formation. In Ref.~\cite{PRC55_Sedrakian}, the onset of superfluidity in asymmetric nuclear matter without like-nucleon correlations was investigated within the BCS framework. It was demonstrated that thermal excitations can actually enable the emergence of a superfluid phase at finite temperature. In contrast, path-integral studies of $pn$ pairing in finite nuclear systems with both like-nucleon and $pn$ correlations, performed for a schematic one-level model~\cite{PRC76_Fellah,IJMPhE27_Mokhtari}  and for even-even $N=Z$ nuclei~\cite{CPC49_Fellah}, found no evidence of enhanced $pn$ pairing at finite temperature.

While earlier studies of $pn$ pairing in hot finite nuclei focused primarily on $N=Z$ systems, the goal of the present work is to explore the emergence of $pn$ pairing reentrance in asymmetric ($N > Z$) even-even nuclei.   We investigate  proton-neutron  pairing correlations in hot nuclei within a finite-temperature generalization of the pnBCS framework.  To this end, we employ the superoperator formalism, which naturally extends zero-temperature many-body concepts, such as vacuum states, quasiparticles, and phonons, etc, to the finite-temperature regime. This formalism has recently been successfully applied to  calculate stellar weak-interaction rates and cross sections for hot nuclei (see, e.g., the review in Refs.~\cite{PhPN53_Dzhioev1,*PhPN53_Dzhioev2,*PhPN53_Dzhioev3} and references therein). Within this framework, we present the  demonstration of temperature-induced reentrance of $pn$ pairing in asymmetric even-even nuclei, explicitly accounting for the competing influence of like-nucleon correlations.  To the best of our knowledge, this is the first study demonstrating $pn$ pairing reentrance in $N > Z$  nuclei with simultaneously treated like-nucleon and $pn$ correlations.
We show that this  reentrance arises from the partial unblocking of single-particle levels that are otherwise Pauli-blocked and suppress $pn$ pair formation at low temperatures. Furthermore, we provide  a  qualitative analysis of how the reentrance of $pn$ pairing modifies the charge-exchange strength distributions in hot nuclear systems.

The paper is organized as follows. Section~\ref{sec:Formalism} presents the application of the superoperator formalism to finite-temperature pairing correlations in nuclear systems. We employ a pairing Hamiltonian that incorporates $pp$, $nn$, and $pn$ monopole interactions in both isovector ($t=1$) and isoscalar ($t=0$) channels, and derive the corresponding finite-temperature gap equations. For completeness, a concise overview of the superoperator formalism is provided in Appendix~\ref{sect:superoperators}.
In Sec.~\ref{sec:results&model} we perform numerical calculations within a schematic equidistant multilevel pairing model to illustrate the non-monotonic temperature dependence of $pn$ pairing and its reentrant enhancement. Section~\ref{sec:results&realistic} demonstrates that this reentrance persists in realistic calculations for even-even Ge isotopes and examines its impact on the Fermi-type charge-exchange strength function. Conclusions and future perspectives are summarized in Sec.~\ref{sec:conclusions}.

\section{FORMALISM \label{sec:Formalism}}

We  investigate pairing correlations in hot even–even atomic nuclei, focusing specifically on a  pairing Hamiltonian that incorporates constant-strength $pp$, $nn$, and $pn$ monopole pairing interactions between nucleons occupying time-reversed single-particle states \cite{vsimkovic2003proton}:
\begin{align}\label{H_pairing}
	&H(a^\dag, a) = \sum_{k\tau} \left(\epsilon_{k\tau} - \lambda_{\tau}\right)(a_{k\tau}^{\dagger} a_{k\tau} +  a_{\bar k\tau}^{\dagger} a_{\bar k\tau}) \\
	&- \sum_\tau G_{\tau}^{(1)} \sum_{k,k'>0} S_{k\tau\tau}^{(1)\dagger} S_{k'\tau\tau}^{(1)}
	- \sum_t G_{pn}^{(t)} \sum_{k,k'>0} S_{kpn}^{(t)\dagger} S_{k' pn}^{(t)}.\notag
\end{align}
 Here \( a_{k\tau}^\dagger \) and \( a_{k\tau} \) denote the creation and annihilation operators for a nucleon in the single-particle state \( k\tau~(\tau =p,\,n)\) with the corresponding single-particle energy \( \epsilon_{k\tau} \). The label \( \bar{k}\tau \) denotes the time-reversed partner of the state \( k\tau \), and \( \lambda_p \) (\( \lambda_n \)) represents the chemical potential introduced as a Lagrange multiplier to adjust the average number of  protons (neutrons) corresponding to the nucleus under consideration.  The operators $S^{(t)\dag}_{k\tau\tau'}$ create isovector ($t=1$) and isoscalar ($t=0$) nucleon pairs in time-reversed states:
\begin{align}
	S_{k\tau\tau}^{(1)\dagger} &= a_{k\tau}^\dagger a_{\bar{k}\tau}^\dagger,  \nonumber \\
	S_{kpn}^{(t)\dagger} &= \frac{1}{\sqrt{2}} (a_{kp}^\dagger a_{\bar{k}n}^\dagger - (-1)^t a_{kn}^\dagger a_{\bar{k}p}^\dagger).
	\end{align}
The Pauli exclusion principle restricts isoscalar pairing exclusively to $pn$ pairs, as identical fermions cannot form $t=0$ states in time-reversed states.
The constant pairing strengths for like nucleons, $G^{(1)}_{n,p}$, and for $pn$ pairs, $G^{(t)}_{pn}$, are taken to be real. In general, we treat $G^{(1)}_p$, $G^{(1)}_n$, $G^{(1)}_{pn}$, and $G^{(0)}_{pn}$ as independent parameters that are allowed to be different.

As briefly outlined in Appendix~\ref{sect:superoperators} (see~\cite{PhPN53_Dzhioev1} for comprehensive treatment), the superoperator formalism provides a  framework for studying quantum many-body systems at finite temperature $T$  by recasting the calculation of equilibrium and non-equilibrium observables into an eigenvalue problem in the Liouville space. For a hot nucleus governed by the pairing Hamiltonian~\eqref{H_pairing}, this reduces to diagonalization of the thermal Hamiltonian \( \mathcal{H} = H - \widetilde{H} \),
where $\widetilde{H}$ is the tilde-conjugate of $H$, contracted by replacing all nucleon creation and annihilation operators with their tilded  counterparts, i.e. $\widetilde{H} = H(\widetilde{a}^\dagger, \widetilde{a})$.  The thermal vacuum $\ket{0(T)}$, the zero-eigenvalue state of the thermal Hamiltonian $\mathcal{H}$ satisfying the thermal state condition (TSC)~\eqref{TSC}, fully characterizes the equilibrium properties of the hot nucleus, i.e. all thermal averages are evaluated as vacuum expectation values, $\bbrakett{X} = \langle 0(T) | X | 0(T) \rangle$. Furthermore, the thermal vacuum serves as a reference vacuum state with respect to which all non-equilibrium excitations induced by external fields are constructed.

Within the independent thermal quasiparticle approximation, the thermal Hamiltonian \( \mathcal{H} \) corresponding to the pairing Hamiltonian~\eqref{H_pairing} is diagonalized in terms of thermal quasiparticle creation and annihilation operators  yielding the canonical form
\begin{equation}\label{H_thermal_diagonal}
     \mathcal{H} \approx \sum_{k,\nu} E_{k\nu} \big( \beta^\dagger_{k\nu} \beta_{k\nu} - \widetilde{\beta}^\dagger_{k\nu} \widetilde{\beta}_{k\nu} \big).
 \end{equation}
The thermal vacuum \( \ket{0(T)} \) is then defined as a vacuum state of thermal quasiparticles (see Eq.~\eqref{annihilation}). Thermal quasiparticles are introduced  through two successive unitary transformations.  The first is the generalized Bogoliubov \((u,v)\)-transformation~\cite{vsimkovic2003proton}, which maps the original nucleon operators onto the quasiparticle operators \( \alpha^\dagger_{k\nu} \) and \( \alpha_{k\nu} \) (\( \nu = 1,2 \)):
\begin{equation}\label{bogo1}
	\begin{pmatrix}
		a_{kp}^{\dagger} \\
		a_{kn}^{\dagger} \\
		a_{\bar{k}p} \\
		a_{\bar{k}n}
	\end{pmatrix}
	 =
	\begin{pmatrix}
		u_{k1p} & u_{k2p} & -v_{k1p} & -v_{k2p} \\
		u_{k1n} & u_{k2n} & -v_{k1n} & -v_{k2n} \\
		v_{k1p} & v_{k2p} & u_{k1p} & u_{k2p} \\
		v_{k1n} & v_{k2n} & u_{k1n} & u_{k2n}
	\end{pmatrix}
	\begin{pmatrix}
		\alpha_{k1}^{\dagger} \\
		\alpha_{k2}^{\dagger} \\
		\alpha_{\bar{k}1} \\
		\alpha_{\bar{k}2}
	\end{pmatrix}.
	\end{equation}
In general, the transformation amplitudes \( u_{k\nu\tau} \) and \( v_{k\nu\tau} \) are complex-valued.
 An analogous transformation with complex-conjugate amplitudes \( u^*_{k\nu\tau} \) and \( v^*_{k\nu\tau} \) is applied to the tilded nucleon operators \( \widetilde{a}^\dagger_{k\tau} \) and \( \widetilde{a}_{k\tau} \) yielding the tilde-quasiparticle operators \( \widetilde{\alpha}^\dagger_{k\nu} \) and \( \widetilde{\alpha}_{k\nu} \). The second, so-called thermal \((x,y)\)-transformation, introduces thermal quasiparticles by mixing  non-tilde and tilde quasiparticle operators
\begin{align}
	\beta_{k \nu}^\dagger &= x_{k \nu} \alpha_{k \nu}^\dagger - i y_{k \nu} \tilde{\alpha}_{k \nu}, \notag\\
	\tilde{\beta}_{k\nu}^\dagger &= x_{k \nu} \tilde{\alpha}_{k \nu}^\dagger + i y_{k \nu} \alpha_{k \nu}.
\end{align}
Here $x_{k \nu}$ and $y_{k \nu}$ are real and $x_{k \nu}^2 + y_{k \nu}^2 = 1$. It should be noted that the complex nature of the thermal transformation arises directly from the TSC~\eqref{TSC} (see Ref.~\cite{PhPN53_Dzhioev1} for a detailed discussion of this point). Since both transformations are unitary, the resulting thermal quasiparticle creation and annihilation superoperators preserve the canonical fermionic anticommutation relations.

To determine the coefficients of the \((u,v)\) and \((x,y)\) transformations, we first observe that if the thermal quasiparticles are eigenmodes of the thermal Hamiltonian  and the vacuum  \( |0(T)\rangle \) of thermal quasiparticles  satisfies the TSC~\eqref{TSC}, then the  amplitudes $y_{k\nu}$ are fixed as
 \begin{equation}\label{y_thermal}
     y_{k\nu} = \left( \frac{1}{1 + e^{E_{k\nu}/T}} \right)^{1/2}.
 \end{equation}
 This result follows directly by inserting, for instance, \( X = \alpha_{k\nu} \) into the TSC~\eqref{TSC}.
On the other hand, $y_{k\nu}^2$ is nothing  but a fermionic occupation number of the quasiparticle state $k\nu$
\begin{equation}
y^2_{k\nu}=\braket{0(T)|\alpha^\dag_{k\nu}\alpha_{k\nu}|0(T)}.
\end{equation}

To determine the amplitudes \((u,v)\), we impose the eigenmode condition
 \begin{equation}\label{EMC}
     [\mathcal{H},\beta^\dagger_{k\nu}] = E_{k\nu}\, \beta^\dagger_{k\nu}.
 \end{equation}
This equation is most conveniently solved by applying Wick’s theorem to bring the thermal Hamiltonian into normal-ordered form with respect to the thermal vacuum. The only non-vanishing contractions are  the following components of the single-particle density matrix \(\rho_{kl}\equiv\bbrakett{a^\dag_k a_l} =\bbrakett{\widetilde a^\dag_k \widetilde a_l}^*\) and paring tensor  \(\kappa_{kl}\equiv\bbrakett{a^\dag_k a^\dag_l}=\bbrakett{\widetilde a^\dag_k \widetilde a^\dag_l}^*\):
\begin{align}\label{def_rho_kappa}
    \rho_{k\tau;k\tau'} &= \sum_\nu\big(y^2_{k\nu}u_{k\nu\tau} u^*_{k\nu\tau'} + x^2_{k\nu}v_{k\nu\tau} v^*_{k\nu\tau'}\big),
    \notag \\
    \kappa_{k\tau;\bar k\tau'} &= \sum_\nu\big(y^2_{k\nu}u_{k\nu\tau} v^*_{k\nu\tau'} - x^2_{k\nu}v_{k\nu \tau} u^*_{k\nu \tau'}\big),
\end{align}
and time-reversal invariance gives  $\rho_{k\tau;k\tau'}^*=   \rho_{\bar k\tau;\bar k\tau'}$, $ \kappa_{k\tau;\bar k\tau'}^*= \kappa_{k\tau';\bar k\tau}$.
 Then, performing linearization, i.e. neglecting in Eq.~\eqref{EMC} the normal product triple terms, we arrive at the eigenvalue problem,  whose solution yields the energy $E_{k \nu}$  and the $(u,v)$-amplitudes of thermal quasiparticles:
\begin{equation}\label{TQBCS}
	\begin{pmatrix}
		\mathcal{E}_{kp} & \mathcal{E}_{kpn} & \Delta_p & \Delta_{pn} \\
		\mathcal{E}_{knp} & \mathcal{E}_{kn} & \Delta_{np} & \Delta_n \\
		\Delta_p & \Delta_{pn} & -\mathcal{E}_{kp} & -\mathcal{E}_{kpn} \\
		\Delta_{np} & \Delta_n & -\mathcal{E}_{knp} & -\mathcal{E}_{kn}
	\end{pmatrix}
	\begin{pmatrix}
		u_{k \nu p} \\
		u_{k \nu n} \\
		v_{k \nu p} \\
		v_{k \nu n}
	\end{pmatrix}
	=E_{k \nu}
	\begin{pmatrix}
		u_{k \nu p} \\
		u_{k \nu n} \\
		v_{k \nu p} \\
		v_{k \nu n}
	\end{pmatrix}.
\end{equation}
Here,  the mean-field potentials  and pairing gaps are given by
\begin{align}\label{BCS_ME}
&\mathcal{E}_{k\tau} = (\epsilon_{k\tau}\!-\!\lambda_\tau)\!-\! G^{(1)}_{\tau}\rho_{k\tau;k\tau}\!-\!\frac12\big(G^{(1)}_{pn}\!+\!G^{(0)}_{pn}\big)\rho_{k\tau';k\tau'},
\notag\\
&\mathcal{E}_{k\tau\tau'} = -\frac12\big(G^{(1)}_{pn}-G^{(0)}_{pn}\big)\rho_{k\tau';k\tau},
\notag\\
&\Delta_\tau = -G^{(1)}_{\tau}\sum_{k>0} \kappa_{k\tau;\bar k\tau},
\\
&\Delta_{\tau\tau'} = -\frac12\sum_{t}\!G^{(t)}_{pn}\!\sum_{k>0} \big(\kappa_{k\tau;\bar k\tau'}\!-\!(-1)^t\kappa_{k\tau';\bar k\tau}\big).\notag
    \end{align}
Hereafter, $\tau'$ denotes the isospin partner of $\tau$ (i.e., $\tau' \neq \tau$).
From the  symmetry properties of the density matrix and the pairing tensor it follows that $\mathcal{E}_\tau$ and $\Delta_\tau$ are real, while $\mathcal{E}_{k\tau\tau'} =\mathcal{E}^*_{k\tau'\tau}$ and $\Delta_{\tau\tau'} =\Delta^*_{\tau'\tau}$.  Consequently, the matrix in~\eqref{TQBCS} is Hermitian, guaranteeing real eigenvalues \( E_{k\nu} \).
 These eigenvalues appear in pairs of opposite signs, a property well known from the zero-temperature pnBCS theory (see, e.g., Ref.~\cite{goswami1964treatment}).
In what follows, the subscript $\nu = 1, 2$ labels the positive-energy solutions $E_{k\nu}$, which correspond to the non-tilde thermal quasiparticles, while their tilde-conjugate partners carry the negative energy $-E_{k\nu}$.

  Using explicit expressions for the pairing tensor $\kappa_{k\tau;\bar k\tau'}$ (see Eq.~\eqref{def_rho_kappa}), we can write the pairing gaps as
\begin{align}\label{gap1}
    &\Delta_\tau = \frac{G^{(1)}_{\tau}}{2}\!\!\sum_{\nu, k>0}(u_{k\nu\tau}v^*_{k\nu\tau}+u^*_{k\nu\tau}v_{k\nu\tau})(1-2y^2_{k\nu}),
    \notag\\
   &\Delta_{\tau\tau'} = \Delta^{(1)}_{\tau\tau'}+ i \Delta^{(0)}_{\tau\tau'},
    \notag\\
   & \Delta^{(1)}_{\tau\tau'} = \frac{G^{(1)}_{pn}}{2}\!\!\sum_{\nu, k>0}\!\!\text{Re}(u_{k\nu\tau}v^*_{k\nu\tau'}\!+\!u_{k\nu\tau'}v^*_{k\nu\tau})(1\!-\!2y^2_{k\nu}),
     \notag\\
   & \Delta^{(0)}_{\tau\tau'} = \frac{G^{(0)}_{pn}}{2}\!\!\sum_{\nu, k>0}\!\!\text{Im}(u_{k\nu\tau}v^*_{k\nu\tau'}\!-\!u_{k\nu\tau'}v^*_{k\nu\tau})(1\!-\!2y^2_{k\nu}).
    \end{align}
 Thus, the real part of the $pn$ pairing gap originates from isovector pairing, whereas its imaginary part stems from isoscalar pairing. This feature, well known from zero-temperature analyses~\cite{chen1967generalized,goodman1968restoration}, persists at finite temperature.

Together with the particle-number constraints
\begin{equation}\label{p_number}
    N_\tau = 2\sum_{k>0} \rho_{k\tau;k\tau}, \quad (\tau = p, n),
\end{equation}
Eqs.~(\ref{y_thermal}), (\ref{TQBCS}), and (\ref{gap1}) form a closed set of finite-temperature pnBCS (FT-pnBCS) equations. Owing to their coupled non-linear structure, these equations must be solved self-consistently. In numerical implementations, an iterative scheme is typically employed, initialized from the zero-temperature pnBCS solution, and then gradually evolved to the desired temperature. Due to their non-linear character, the FT-pnBCS equations may admit multiple stationary solutions, depending on the initial zero-temperature seed. The physically relevant solution is uniquely identified as the one that minimizes the free energy.

 Analytical expressions for the quasiparticle energies \( E_{k\nu} \) can be derived by employing the reduced form of the squared FT-pnBCS matrix, following the method developed for the zero-temperature case~\cite{goswami1964treatment}. These expressions are considerably simplified under the approximation that single-particle states are not renormalized, specifically, by setting \( \mathcal{E}_{k\tau} = \epsilon_{k\tau}-\lambda_\tau \) and \( \mathcal{E}_{k\tau\tau'} = 0 \). Namely,
\begin{equation}\label{eq:E_knu}
    E_{k\nu}^2=\frac12\big(E^2_{kp}\!+\!E^2_{kn}\!+\!2|\Delta_{pn}|^2\!+\!(-1)^{\nu+1}R_{k}\Big),~~\nu=1,\,2,
\end{equation}
where $E_{k\tau}=\sqrt{\varepsilon_{k\tau}^2+\Delta_\tau^2}$ ($\varepsilon_{k\tau}=\epsilon_{k\tau}-\lambda_\tau$) are conventional BCS quasiparticle energies, and
\[
R_{k}\!=\!\Big[(E^2_{kp}-E^2_{kn})^2+4|\Delta_{pn}|^2\big((\Delta_p+\Delta_n)^2+(\varepsilon_{kp}-\varepsilon_{kn})^2\big)\Big]^{1/2}.
\]
As it should be  $E_{k1(2)}=E_{kp(n)}$ for vanishing proton-neutron pairing.

Furthermore, within the above approximation, analytical expressions for the $(u,v)$-coefficients can be derived. Substituting these expressions into Eqs.~\eqref{gap1} and~\eqref{p_number} then yields a coupled system of FT-pnBCS gap equations:
\begin{align}\label{gap2}
&N_\tau=\sum_{k>0}\bigg[1\!-\!\sum_{\nu}\tanh\Big(\frac{E_{k\nu}}{2T}\Big)\frac{1}{2E_{k\nu}}
\notag\\
&\!\!\times\!\!\bigg\{\!\varepsilon_{k\tau}\!+\!(-1)^{1+\nu}\frac{(E^2_{k\tau}\!-\!E^2_{k\tau'})\varepsilon_{k\tau}\!+\!2 |\Delta_{pn}|^2(\varepsilon_{k\tau}\!-\!\varepsilon_{k\tau'})}{R_k}\!\!\bigg\}\!\bigg]\!,
\notag\\
   &\frac{4\Delta_{\tau}}{G^{(1)}_{\tau}}=\sum_{\nu,k>0}\tanh\Big(\frac{E_{k\nu}}{2T}\Big)\frac{1}{E_{k\nu}}
   \notag\\
   &\times\bigg\{\!\Delta_\tau\!+\!(-1)^{1+\nu}\frac{(E^2_{k\tau}\!-\!E^2_{k\tau'})\Delta_\tau\! +\!2|\Delta_{pn}|^2(\Delta_p\!+\!\Delta_n)}{R_k}\!\bigg\},
   \notag\\
   &\frac{4\Delta^{(t)}_{pn}}{G^{(t)}_{pn}}=\Delta^{(t)}_{pn}\sum_{\nu,k>0}\tanh\Big(\frac{E_{k\nu}}{2T}\Big)\frac{1}{E_{k\nu}}
   \notag\\
   &\times\bigg\{1\!+\!(-1)^{1+\nu}\frac{E^2_{kp}\!+\!E^2_{kn} +2 (\Delta_p\Delta_n\!-\!\varepsilon_{kp}\varepsilon_{kn})}{R_k}\bigg\},
\end{align}
where $\tanh\Big(\frac{E_{k\nu}}{2T}\Big)=1-2y^2_{k\nu}$. The same  gap equations were derived in~\cite{IJMPhE27_Mokhtari}
 using a path-integral approach.

The key feature of the FT-pnBCS gap equations~\eqref{gap2} is that the summation term in the last equation  is identical for both the $t=0$ and $t=1$ isospin channels. Consequently, when $G^{(1)}_{pn} \neq G^{(0)}_{pn}$, the isovector and isoscalar $pn$ pairing correlations cannot coexist. At zero temperature,  this result was observed in~\cite{PLB393_Satula,vsimkovic2003proton} by numerical solution of the pnBCS gap equations.   For $G_{pn}^{(t)}>G^{(t')}_{pn}$,  the solution that minimizes the free energy is $\Delta_{pn}^{(t)}\ne 0$ and  $\Delta_{pn}^{(t')}=0$. In the particular case  $G^{(1)}_{pn}= G^{(0)}_{pn}$, both $t=1$ and $t=0$ pairing modes coexist and $|\Delta^{(t)}_{pn}| = |\Delta_{pn}|/\sqrt{2}$. Furthermore, since the gap equations~\eqref{gap2} are invariant under the transformation $\Delta^{(t)}_{pn} \to -\Delta^{(t)}_{pn}$, the sign of $\Delta^{(t)}_{pn}$ is arbitrary and we adopt the convention $\Delta^{(t)}_{pn} \ge 0$ without loss of generality. In contrast, the relative
sign of $\Delta_p$ and $\Delta_n$ has physical significance when $\Delta_{pn}\ne 0$,  as it directly influences the coupling between like-nucleon and $pn$ pairing channels.

 Assuming $\Delta_{pn}=0$ in~\eqref{gap2}, we  obtain a critical value of the $pn$ interaction strength below which $pn$-pairing correlations vanish
\begin{align}\label{eq:Gpn_cr}
\frac{2}{G^\text{cr}_{pn}}=\sum_{\tau,k>0}\tanh\Big(\frac{E_{k\tau}}{2T}\Big)\frac{E^2_{k\tau}+\Delta_p\Delta_n -\varepsilon_{kp}\varepsilon_{kn}}{E_{k\tau}(E^2_{k\tau}-E^2_{k\tau'})}.
\end{align}
In the next section, we will demonstrate that \( G^{\text{cr}}_{pn} \) can exhibit a non-monotonic dependence on temperature. This, in turn, leads to an enhancement of $pn$ pairing correlations in certain temperature regimes.

A further simplification of the gap equations~\eqref{gap2} is possible for self-conjugated nuclei ($N_n=N_p$) when the Coulomb interaction is neglected. Under these conditions $\varepsilon_{kp}=\varepsilon_{kn}$ and $G^{(1)}_{p}=G^{(1)}_{n}$. Moreover, charge symmetry requires that the pairing tensors satisfy $ \kappa_{kp;\bar kp}= -\kappa_{kn;\bar k n}$~\cite{NPA167_Wolter}. Therefore, $\Delta_n=-\Delta_p$, $R_k=0$ and~\eqref{gap2} reduces to a simple form
\begin{align}\label{eq:gap3}
&N_\tau=\sum_{\nu,k>0}\bigg[1-\tanh\Big(\frac{E_{k\nu}}{2T}\Big)\frac{\varepsilon_{k\tau}}{E_{k\nu}}\bigg],
\notag\\
 &\frac{4\Delta_{\tau}}{G^{(1)}_{\tau\tau}}=\Delta_{\tau}\sum_{\nu,k>0}\tanh\Big(\frac{E_{k\nu}}{2T}\Big)\frac{1}{E_{k\nu}},
   \notag\\
 &\frac{4\Delta^{(t)}_{pn}}{G^{(t)}_{pn}}=\Delta^{(t)}_{pn}\sum_{\nu,k>0}\tanh\Big(\frac{E_{k\nu}}{2T}\Big)\frac{1}{E_{k\nu}},
\end{align}
where the  quasiparticle energies $E_{k\nu} = (\varepsilon_{k\tau}^2 + \Delta_\tau^2 + |\Delta_{pn}|^2)^{1/2}$ are twice degenerate.
Thus, in self-conjugated nuclei, like-nucleon and $pn$ pairing correlations do not coexist within the FT-pnBCS approximation unless the isovector strength $G^{(1)}_{n,p}$ equals the $pn$ interaction strength in either the isoscalar or isovector channel. Isospin symmetry requires the isovector pairing strengths to be equal, i.e., $G^{(1)}_{n,p}=G^{(1)}_{pn}$. Under this condition, isovector pairing dominates when $G^{(1)}_{n,p} > G^{(0)}_{pn}$, whereas isoscalar pairing becomes energetically favored for $G^{(0)}_{pn} > G^{(1)}_{n,p}$.
Importantly,  in the isospin-symmetric limit with $G^{(0)}_{pn} > G^{(1)}_{n,p}$, isovector like-nucleon and isoscalar $pn$ pairing may coexist within the approximate Lipkin-Nogami particle-number projection scheme~\cite{PLB393_Satula}.

For completeness, we present explicit expressions for the total energy and entropy used to evaluate the free energy, $F=E-TS$, within the FT-pnBCS approximation:
\begin{align}\label{energy}
  &  E  = \braket{0(T)|H|0(T)} + \sum_\tau \lambda_\tau N_\tau
    \notag \\
  &  =\sum_{k>0}(\epsilon_{pk}+\epsilon_{nk})-\sum_{\nu,k>0}\tanh\Big(\frac{E_{k\nu}}{2T}\Big)\frac{1}{2E_{k\nu}}
    \notag\\
  &  \times\Big\{(\epsilon_{pk}\varepsilon_{pk}+\epsilon_{nk}\varepsilon_{nk})+\frac{(-1)^{1+\nu}}{R_k}\big[(\epsilon_{pk}\varepsilon_{pk}-\epsilon_{nk}\varepsilon_{nk})
    \notag\\
  &  \times(E^2_{kp}-E^2_{kn})+2|\Delta_{pn}|^2(\epsilon_{pk}-\epsilon_{nk})(\varepsilon_{pk}-\varepsilon_{nk})\big]\Big\}
  \notag\\
 & -\frac{\Delta_n^2}{G^{(1)}_n}-\frac{\Delta_p^2}{G^{(1)}_p}-2\frac{(\Delta^{(0)}_{pn})^2}{G^{(0)}_{pn}}-2\frac{(\Delta^{(1)}_{pn})^2}{G^{(1)}_{pn}},
\end{align}
\begin{align}\label{entropy}
    S = -2\sum_{\nu,k>0} \big[ y^2_{k\nu}\ln{y^2_{k\nu}} + (1-y^2_{k\nu})\ln{(1-y^2_{k\nu}})\big].
\end{align}
In the absence of $pn$ pairing, these expressions reduce to the corresponding FT-BCS results given in Ref.~\cite{NPA182_Moretto}.

\section{NUMERICAL CALCULATIONS AND RESULTS}

In this section, we apply the formalism developed above to study $pn$ pairing in hot nuclear systems, including a schematic model with equidistant doubly degenerate single-particle levels and realistic even–even germanium isotopes. Prior to presenting the results, we first outline the key technical aspects of the numerical implementation.

To solve the highly nonlinear coupled gap equations~\eqref{gap2}, we employed the NEQBF subroutine from the IMSL Fortran Numerical Library~\cite{IMSL}, which solves a system of nonlinear equations using factored secant update with a finite-difference approximation to the Jacobian. As was mentioned above, the iterative solution procedure is initialized with the zero-temperature pnBCS solution and then continued to the target temperature by incrementally increasing
$T$ and using the solution at each step as the initial guess for the next.
This continuation strategy ensures convergence across the full temperature range.  Depending on the initial zero-temperature seed, the FT-pnBCS equations may admit multiple stationary solutions.  A crucial numerical aspect is that a non-trivial $pn$ pairing solution ($\Delta_{pn} \neq 0$) is obtained only when the iterative procedure is initialized with like-nucleon gaps of opposite sign, i.e., $\Delta_p \cdot \Delta_n < 0$. This sign asymmetry in the initial conditions reflects the properties of charge conjugation operation~\cite{NPA167_Wolter}, which leads to $\Delta_p=-\Delta_n$ in isospin symmetric  nuclei. For clarity, all subsequent figures and discussions display the absolute values $|\Delta_p|$, $|\Delta_n|$, and we adopt $\Delta^{(t)}_{pn} \ge 0$ by convention.  To further verify that the $pn$ pairing solution  indeed corresponds to  the global minimum of the free energy (and not to a metastable branch selected by the continuation procedure), we performed two  independent checks. First, we initialized the iterative solver at several intermediate temperatures with different starting values  for the pairing gaps $\Delta_p$, $\Delta_n$, and $\Delta_{pn}$; in all cases, the iterative procedure converged to the same solution obtained by the  continuation strategy. Second,  we compared the free energies of the three competing stationary  solutions, as illustrated in Fig.~\ref{fig:Free_energy} below.

As demonstrated above, within the  FT-pnBCS framework, the $t=1$ and $t=0$ $pn$ pairing channels
are mutually exclusive when the coupling strength $G^{(1)}_{pn}$ and $ G^{(0)}_{pn}$ are unequal. The system instead selects the channel with the stronger interaction,  as it minimizes the free energy.
Consequently, for the numerical calculations presented in this section we restrict ourselves to a single isospin channel, either pure isovector or pure isoscalar, and denote the corresponding pairing strength and gap simply by $G_{pn}$ and $\Delta_{pn}$.

Moreover, we restrict our analysis to asymmetric systems ($N_n > N_p$). In the symmetric limit (see Eqs.~\eqref{eq:gap3}), whenever a finite $pn$ gap $\Delta^{(t)}_{pn}$ exists, its thermal evolution parallels  that of the like-nucleon gaps $\Delta_{p,n}$:   $\Delta^{(t)}_{pn}$ decreases monotonically with temperature and vanishes at a critical temperature, exhibiting no reentrant behavior.

\subsection{Schematic model \label{sec:results&model}}

As a schematic model, we consider a system with $\Omega=10$ doubly degenerate equidistant single-particle levels for both protons and neutrons. The single-particle energies $\epsilon_k$ ($k=1,\dots,\Omega$) are identical in the proton and neutron subsystems and increase linearly with $k$, with a constant spacing $\epsilon_{k+1}-\epsilon_k = 1$ set as the energy unit for all quantities in the calculation (pairing gaps, temperature,  etc.). The double degeneracy of each level accounts for time-reversed partners $k$ and $\bar k$. The proton number is fixed at $N_p = 10$, while the neutron number $N_n$, the like-nucleon pairing strengths $G_{n,p}$, and the $pn$ pairing strength $G_{pn}$ are treated as variable  parameters.

\begin{figure}[h]
	\centering
	\includegraphics[width=\columnwidth]{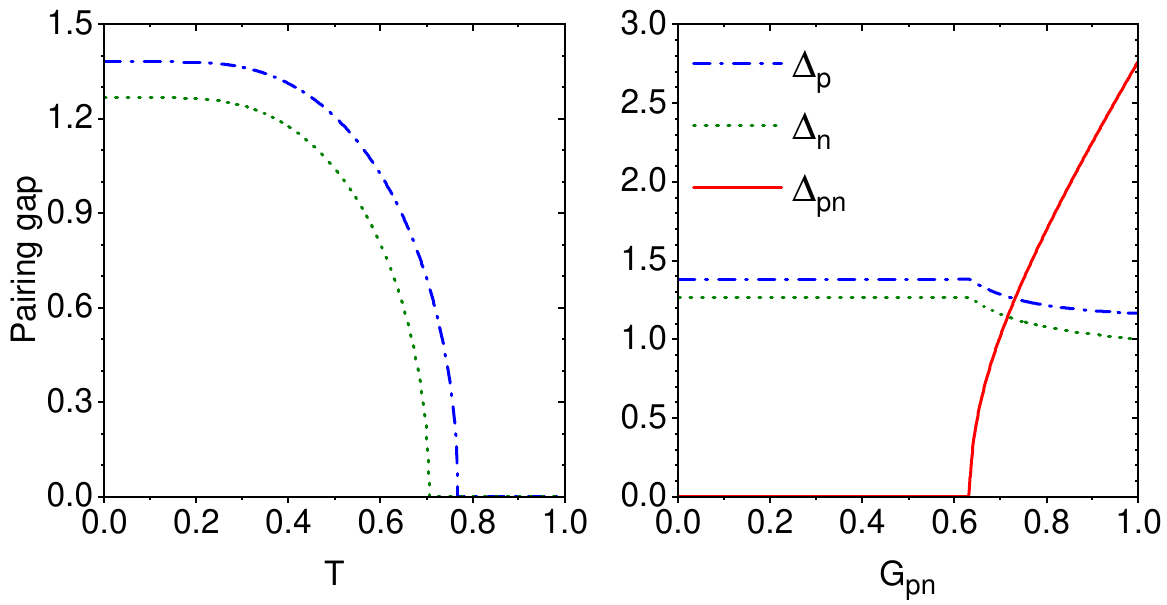}
	\caption{(a) Temperature dependence of the proton ($\Delta_p$) and neutron ($\Delta_n$) pairing gaps in the absence of proton-neutron pairing ($G_{pn}=0$). (b) Ground-state ($T=0$) pairing gaps $\Delta_p$, $\Delta_n$, and $\Delta_{pn}$ as functions of the proton-neutron pairing strength $G_{pn}$. In both panels the particle numbers are fixed at $N_p=10$ and $N_n=14$, and the like-nucleon pairing strength is $G_{n,p}=0.5$.}
	\label{fig:Pairing_gap}
\end{figure}

Figure~\ref{fig:Pairing_gap}(a) displays the temperature dependence of the proton and neutron pairing gaps for a system with  $N_n=14$ and $G_{n,p}=0.5$ calculated in the absence of proton–neutron pairing ($G_{pn}=0$). As expected within the finite-temperature BCS framework, both gaps decrease monotonically with increasing temperature and vanish at their respective critical temperatures.  Since the proton and neutron pairing strengths are equal  and the proton subsystem is exactly half-filled, the proton gap remains larger than the neutron gap at all temperatures and its critical temperature is higher.

Keeping all other parameters fixed, Fig.~\ref{fig:Pairing_gap}(b) illustrates the onset of $pn$ pairing correlations in the ground state ($T=0$) as the interaction strength $G_{pn}$  increases. A critical value $G_{pn}^{\mathrm{cr}}$ exists below which $\Delta_{pn}=0$. For $G_{pn}>G_{pn}^{\mathrm{cr}}$, $\Delta_{pn}$ becomes finite and grows monotonically with $G_{pn}$. The like-nucleon gaps $\Delta_p$ and $\Delta_n$ remain unchanged for $G_{pn}\le G_{pn}^{\mathrm{cr}}$ and decrease only slightly once $pn$ pairing sets in. This behavior is characteristic of the pnBCS approach with a constant pairing interaction: the emergence of  proton-neutron correlations competes weakly with like-nucleon pairing, allowing both types of correlations to coexist above the critical strength -- a feature previously reported in the literature \cite{vsimkovic2003proton,civitarese1997neutron}.

\begin{figure}[h]
	\centering
	\includegraphics[width=\columnwidth]{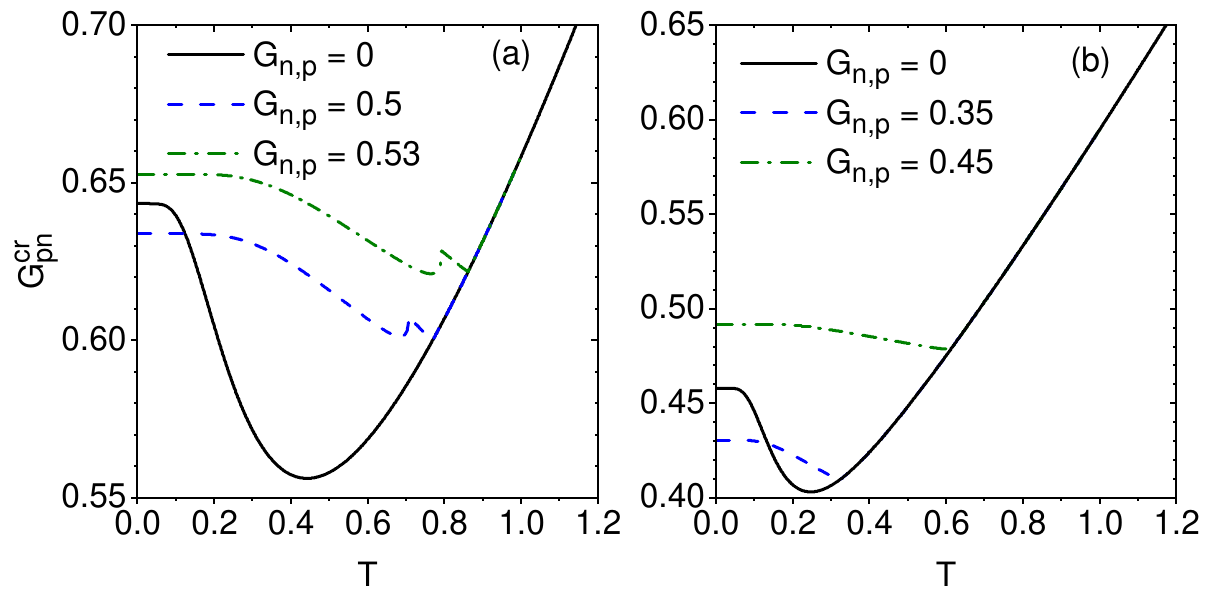}
	\caption{Critical $pn$ pairing strength $G_{pn}^{\mathrm{cr}}$ as a function of temperature $T$ for several values of the like-nucleon pairing strengths $G_{n,p}$. Panel (a) corresponds to $N_p=10$, $N_n=14$; panel (b) shows the case $N_p=10$, $N_n=12$. }
	\label{fig:Gpn_critical}
\end{figure}

The key question is how the critical strength $G_{pn}^{\mathrm{cr}}$ for the onset of $pn$ pairing depends on temperature and  preexisting like-nucleon pairing correlations. To address this, Fig.~\ref{fig:Gpn_critical}(a) displays the temperature dependence of $G_{pn}^{\mathrm{cr}}$ for the $N_p=10$, $N_n=14$ system computed using Eq.~\eqref{eq:Gpn_cr} for two scenarios: with and without like-nucleon pairing included.
In both scenarios, $G_{pn}^{\mathrm{cr}}$ exhibits  non-monotonic behavior: it initially decreases with rising temperature, reaches a minimum at some intermediate temperature, and subsequently increases at higher temperatures.
To demonstrate that the observed non-monotonic behavior is not an artifact of particular proton–neutron asymmetry, Fig.~\ref{fig:Gpn_critical}(b) presents the temperature dependence of $G_{pn}^{\mathrm{cr}}$ for an alternative configuration with $N_n = 12$ and $N_p = 10$.
The qualitative similarity with the $N_n = 14$ case suggests that the observed non-monotonic temperature dependence of $G_{pn}^{\mathrm{cr}}$ is not limited to a single choice of particle numbers or asymmetry.

It is also evident that the increase in $G_{pn}^{\mathrm{cr}}$ observed in Fig.~\ref{fig:Gpn_critical} with increasing neutron number reflects the Pauli blocking effect induced by the neutron excess above the $N_p=N_n$ core: excess neutrons occupy single-particle levels that would otherwise be available for proton-neutron pairing (see Fig.~2 in~\cite{satula2000number} for a schematic illustration of the blocking mechanism for $pn$ pairing). Furthermore, the non-monotonic dependence of $G_{pn}^{\mathrm{cr}}$   on like-nucleon pairing strengths and temperature demonstrates that both thermal excitations and like-nucleon correlations can either suppress or enhance  $pn$ pairing, depending on the specific conditions.  Since like-nucleon pairing is itself temperature-dependent, these two mechanisms are intrinsically coupled. Consequently, the formation of proton-neutron pairing correlations in hot nuclei is governed by a delicate interplay between thermal effects and like-nucleon pairing.

\begin{figure}[h]
	\centering
	\includegraphics[width=\columnwidth]{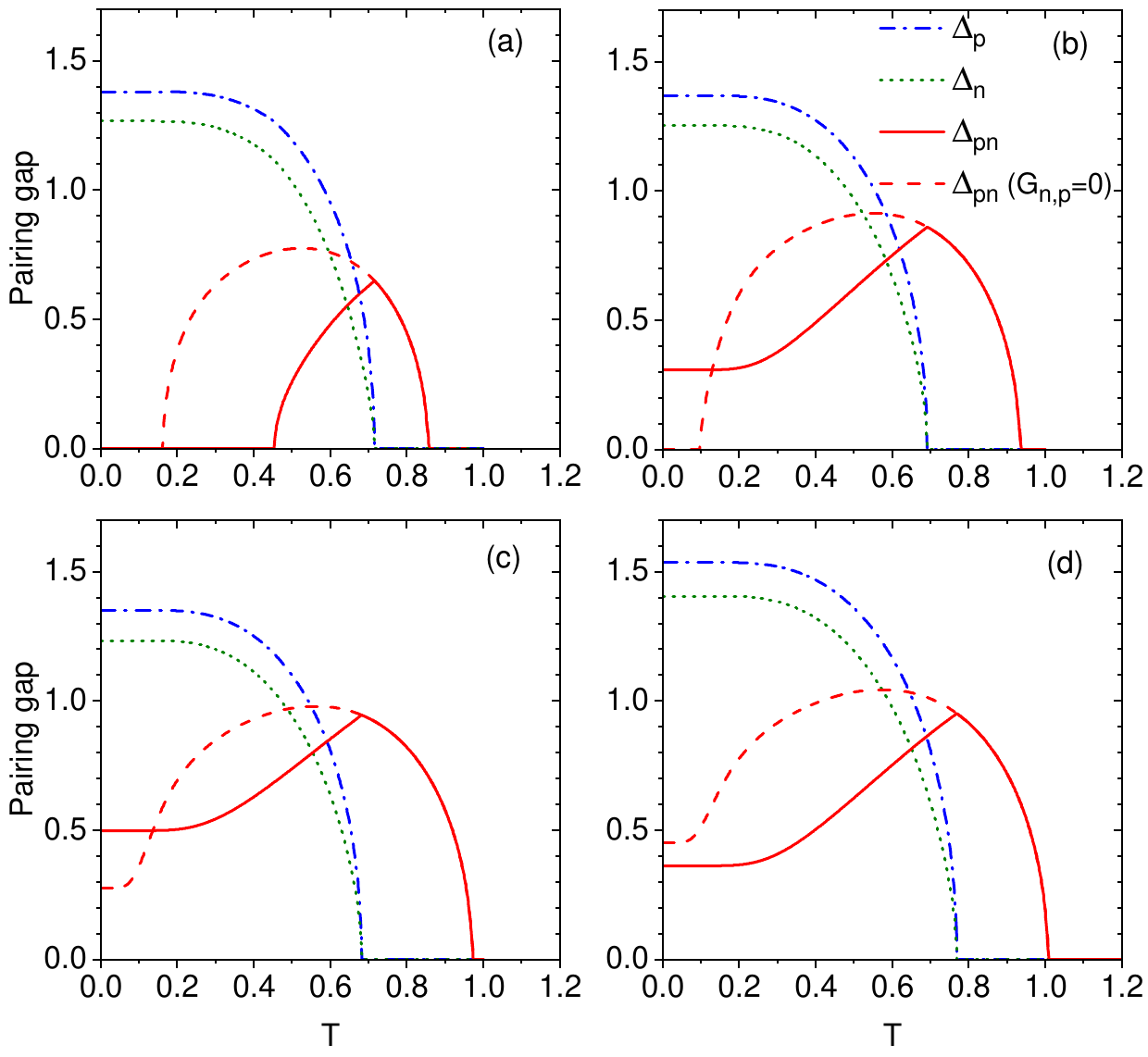}
	\caption{Temperature dependence of the proton, neutron, and $pn$  pairing gaps for selected combinations of the interaction strengths $G_{n,p}$ and $G_{pn}$. Panels (a)--(c) correspond to $G_{n,p}=0.50$ with $G_{pn}=0.62$, $0.64$, and $0.65$, respectively; panel (d) shows the case $G_{n,p}=0.53$, $G_{pn}=0.66$. The particle numbers ($N_p=10$, $N_n=14$) and the number of doubly degenerate single-particle levels ($\Omega=10$) are kept fixed. The reentrance of $\Delta_{pn}$ at intermediate temperatures, coinciding with the quenching of like-nucleon pairing, is clearly visible in all panels.}
	\label{fig:Pairing_gaps_T}
\end{figure}

This interplay is illustrated in Fig.~\ref{fig:Pairing_gaps_T}, which displays the temperature evolution of the neutron, proton, and $pn$ pairing gaps for several combinations of the interaction strengths $G_{pn}$ and $G_{n,p}$.  The figure reveals how the competition between thermal effects and the effect of like-nucleon pairing correlations governs the non‑monotonic behavior of $\Delta_{pn}$, including its characteristic reentrance at intermediate temperatures when like-nucleon pairing is quenched. This behavior of $pn$ pairing is analogous to the reentrance phenomenon of like-nucleon pairing, which appears in hot odd nuclei \cite{HungPRC2016,dong2023pairing} and/or hot rotating nuclei \cite{HungPRC2008,HungPRC2011,HungROPP}.  Notably, in contrast to the $G_{pn}=0$ case shown in Fig.~\ref{fig:Pairing_gap}(a), the critical temperatures, at which the proton and neutron pairing gaps vanish, coincide when $pn$ pairing is present. This synchronization arises because $pn$ correlations couple the proton and neutron subsystems, forcing their pairing phase transitions to occur simultaneously despite the underlying particle-number asymmetry.

To isolate the thermal unblocking mechanism from the influence of like-nucleon pairing correlations, Fig.~\ref{fig:Pairing_gaps_T} additionally displays the temperature dependence of $\Delta_{pn}$ in the limiting case $G_{n,p}=0$. This reference calculation reveals how thermal effects alone govern the onset and evolution of $pn$ pairing correlations.  First, we observe that at moderate temperatures and in the absence of like-nucleon pairing, $\Delta_{pn}$ evolves more rapidly with temperature and attains its maximum value at lower temperatures compared to calculations that include like-nucleon correlations. This comparison demonstrates that  like-nucleon pairing suppresses $pn$ correlations at moderate temperatures.
In contrast, at low temperatures the influence of like-nucleon pairing on the $pn$ gap is ambivalent: depending on the magnitude of $G_{n,p}$, like-nucleon correlations can either enhance or suppress $\Delta_{pn}$.

\begin{figure}[h]
	\centering
	\includegraphics[width=0.8\columnwidth]{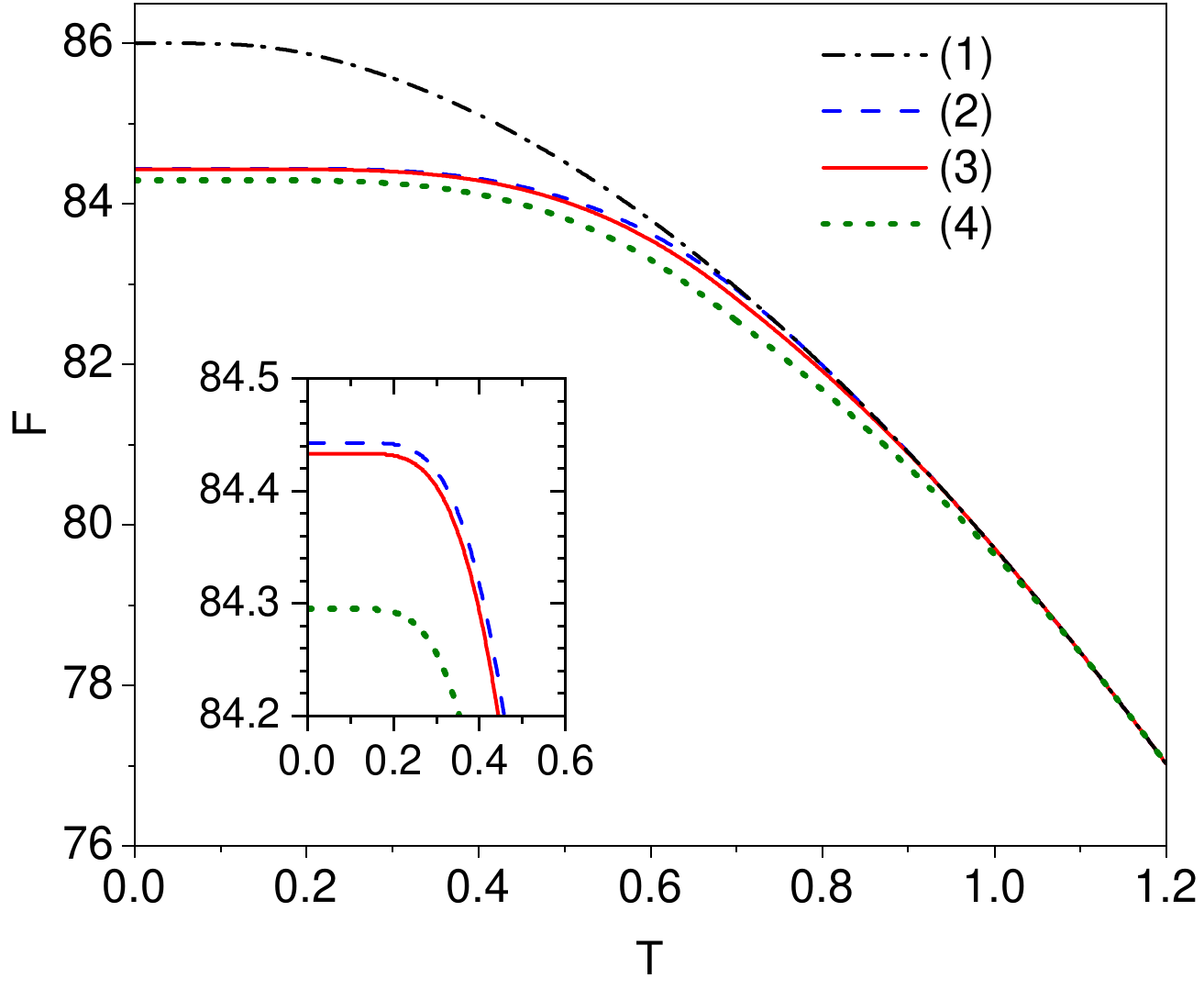}
	\caption{Free energy as a function of temperature for the parameter set of Fig.~\ref{fig:Pairing_gaps_T}(c), evaluated for three distinct solutions of the gap equations~\eqref{gap2}: (1) the trivial normal-phase solution; (2) the solution with only like-nucleon pairing; and (3) the  solution with all pairing channels active. Solution (4) corresponds to  $G_{pn}=0.7$. }
	\label{fig:Free_energy}
\end{figure}

 As mentioned above, the highly nonlinear coupled gap equations~\eqref{gap2} may admit multiple stationary solutions. To identify the physically relevant state, we select the solution that minimizes the free energy $F = E - TS$, computed within the FT-pnBCS approximation (see Eqs.~(\ref{energy},\,\ref{entropy})). As an illustrative example, Fig.~\ref{fig:Free_energy} displays the free energy evaluated for the three distinct solutions of Eqs.~\eqref{gap2}, using the same parameter set as in Fig.~\ref{fig:Pairing_gaps_T}(c). As expected, the onset of nonzero like-nucleon pairing significantly lowers the free energy compared to the normal phase. The difference in free energy between the two superfluid branches (with and without $pn$ pairing) is relatively modest, which is natural since the chosen interaction strength $G_{pn}=0.65$ lies only slightly above the critical value for the onset of $pn$ pairing. Nevertheless, as clearly evident from the figure, the  reentrant $pn$ pairing solution consistently corresponds to the  minimum of the free energy throughout the entire temperature range where it exists. The figure also illustrates the decrease in the free energy with increasing $G_{pn}$.

In the limiting case $G_{n,p}=0$, the FT-pnBCS equations admit an analytical reduction. Specifically, the quantity $R_k$ defined in Eq.~\eqref{eq:E_knu} takes the form
\begin{equation}
  R_k = 2|\lambda_p\! -\! \lambda_n|\,C_k,\quad C_k\!=\!\bigg[\big(\epsilon_k\! -\!\frac12(\lambda_p+\lambda_n)\big)^2+\Delta_{pn}^2\bigg]^{1/2},
\end{equation}
and the quasiparticle energies become
\begin{equation}
  E_{k1} = E_k^+, \qquad E_{k2} = |E_k^-|,
\end{equation}
where $E_k^\pm = C_k \pm \frac{1}{2}|\lambda_p - \lambda_n|$. Note that while $E_k^+$ is always positive, $E_k^-$ becomes negative for single-particle levels satisfying $(\epsilon_k - \lambda_p)(\epsilon_k - \lambda_n) < -\Delta_{pn}^2$. All such levels lie between the proton and neutron chemical potentials, where the product $(\epsilon_k - \lambda_p)(\epsilon_k - \lambda_n)$ is negative.
Substituting these expressions for $R_k$ and $E_{k\nu}$ into the gap equation~\eqref{gap2} yields a compact form for the $pn$ pairing gap:
\begin{equation}
\frac{4\Delta_{pn}}{G_{pn}}\! =\! \Delta_{pn}\!\sum_k \frac{1}{C_k}\left\{\tanh\left(\frac{E_{k1}}{2T}\right)\! +\! \frac{E_k^-}{E_{k2}}\,\tanh\left(\frac{E_{k2}}{2T}\right)\right\},
\end{equation}
or, equivalently,
\begin{equation}\label{eq:gap4}
\frac{2\Delta_{pn}}{G_{pn}} = \Delta_{pn}\sum_k \frac{1}{C_k}\bigl[1 - f(E_k^+) - f(E_k^-)\bigr],
\end{equation}
where $f(E) = \bigl(1 + e^{E/T}\bigr)^{-1}$ is the Fermi–Dirac distribution function. Equation~\eqref{eq:gap4} represents the finite-temperature generalization of the $pn$ pairing gap equation originally derived by Soloviev using Green's function techniques~\cite{Soloviev1963}. A closely related gap equation for asymmetric nuclear matter was subsequently obtained by Sedrakian \textit{et al.} within a similar formalism~\cite{PRC55_Sedrakian}.

The simplified gap equation~\eqref{eq:gap4} reveals that, in the absence of like-nucleon pairing, thermal effects enter exclusively through the occupation factors $f(E_k^\pm)$. Notably, for single-particle levels with energies lying between the proton and neutron chemical potentials, the combination $1 - f(E_k^+) - f(E_k^-)$ acts as an effective blocking factor at low temperatures. In this energy window, where $E_k^- < 0$, the two Fermi–Dirac terms nearly cancel the unity, strongly suppressing the contribution $1/C_k$ to the gap equation. As temperature increases, thermal smearing reduces this cancellation, thereby unblocking these levels and enhancing the phase space available for $pn$ pair formation -- a mechanism that underlies the  reentrance of $\Delta_{pn}$.

\begin{figure}[h]
	\centering
	\includegraphics[width=\columnwidth]{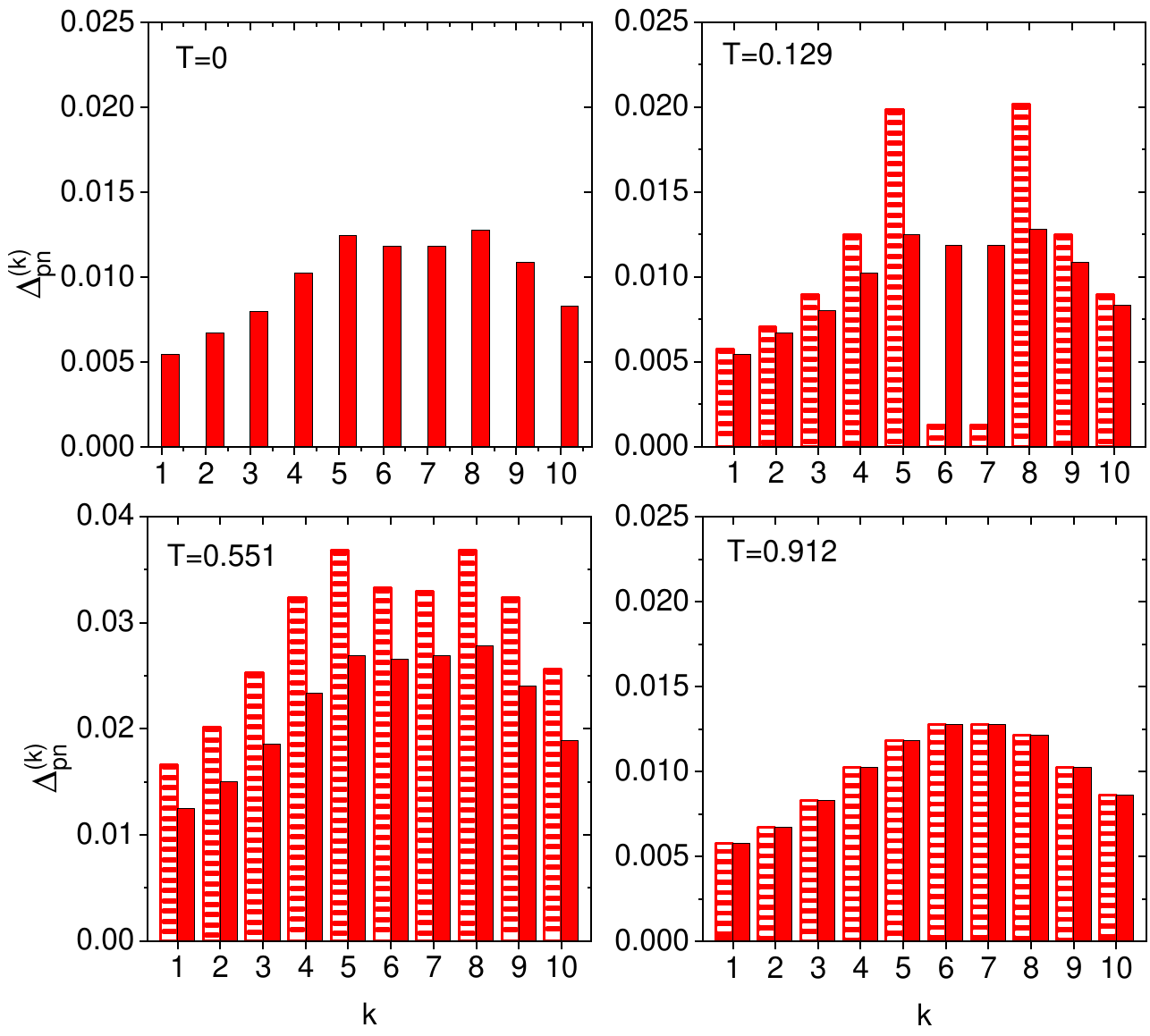}
	\caption{Level-resolved contributions $\Delta_{pn}^{(k)}$ to the proton-neutron pairing gap for the parameter set of Fig.~\ref{fig:Pairing_gaps_T}(b). Solid bars show the results with like-nucleon pairing present ($G_{n,p}=0.5$); dashed bars show the results without like-nucleon pairing ($G_{n,p}=0$). The results are shown at four characteristic temperatures illustrating the evolution of thermal unblocking (see main text). }
	\label{fig:Delta_PN_k}
\end{figure}

The progressive unblocking of single-particle levels is directly reflected in the temperature evolution of the level-resolved  contributions $\Delta_{pn}^{(k)}$ to the $pn$ pairing gap defined by $\Delta_{pn} = \sum_{k=1}^{\Omega} \Delta_{pn}^{(k)}$. Tracking the redistribution of $\Delta_{pn}^{(k)}$ across the single-particle spectrum with increasing temperature reveals which levels become accessible for $pn$ pair formation and identifies the spectral regions where pairing correlations are most concentrated.
 Figure~\ref{fig:Delta_PN_k} displays the contributions $\Delta_{pn}^{(k)}$ for the same parameter set used in Fig.~\ref{fig:Pairing_gaps_T}(b).
 Comparative calculations with and without like-nucleon pairing reveal how thermal unblocking and the interplay with $pp$/$nn$ correlations reshape the spectral distribution of $pn$ pairing as a function of temperature.
  To elucidate the unblocking mechanism, the contributions $\Delta_{pn}^{(k)}$ are displayed at four characteristic temperatures corresponding to the evolution shown in Fig.~\ref{fig:Pairing_gaps_T}(b): (i) $T=0$, where $pn$ pairing is absent in the $G_{n,p}=0$ case; (ii) $T=0.129$, where $\Delta_{pn}$ takes identical values for $G_{n,p}=0.5$ and $G_{n,p}=0$; (iii) $T=0.551$, where thermal unblocking maximizes $\Delta_{pn}$ in the absence of like-nucleon pairing; and (iv) $T=0.912$, where $\Delta_{pn}$ returns to its $T=0.129$ value but with like-nucleon correlations fully quenched.

As shown in Fig.~\ref{fig:Delta_PN_k}, the level-resolved contributions $\Delta^{(k)}_{pn}$  differ markedly in calculations with and without like-nucleon pairing at low temperatures. When like-nucleon pairing is present, it modifies the single-particle occupation factors, effectively opening additional channels for $pn$ pairing.  As a result, the contribution of levels lying between the proton and neutron chemical potentials, predominantly $k=6$ and $k=7$ for the present particle numbers, is  already significant at $T=0$.
In contrast, in the absence of like-nucleon correlations, these same levels contribute only weakly to $\Delta_{pn}$ at low temperatures. Instead, thermal excitations must first unblock levels adjacent to the chemical potential window ($k\neq 6,7$), allowing $pn$ pairing to develop there before eventually migrating toward $k=6,7$ as temperature increases.

The dual role of like-nucleon pairing becomes evident when comparing the distribution of $\Delta_{pn}^{(k)}$ at $T=0$ and $T=0.551$. The observed temperature-dependent crossover from constructive to destructive influence highlights the delicate balance between like-nucleon pairing and thermal effects.
Furthermore, the $T=0.551$ plot reveals that the finite-temperature reentrance of $pn$ pairing involves contributions from all single-particle levels, although the dominant share of $\Delta_{pn}$ originates from levels in the vicinity of the chemical potentials. Intriguingly, at $T=0.912$ the distribution $\Delta_{pn}^{(k)}$ closely resembles that at $T=0$, despite the fact that the underlying unblocking mechanisms are fundamentally different: at $T=0$ the enhancement stems from the occupation smearing induced by like-nucleon pairing, whereas at $T=0.912$ it results from the thermal smearing.

\begin{figure}[h]
	\centering
	\includegraphics[width=\columnwidth]{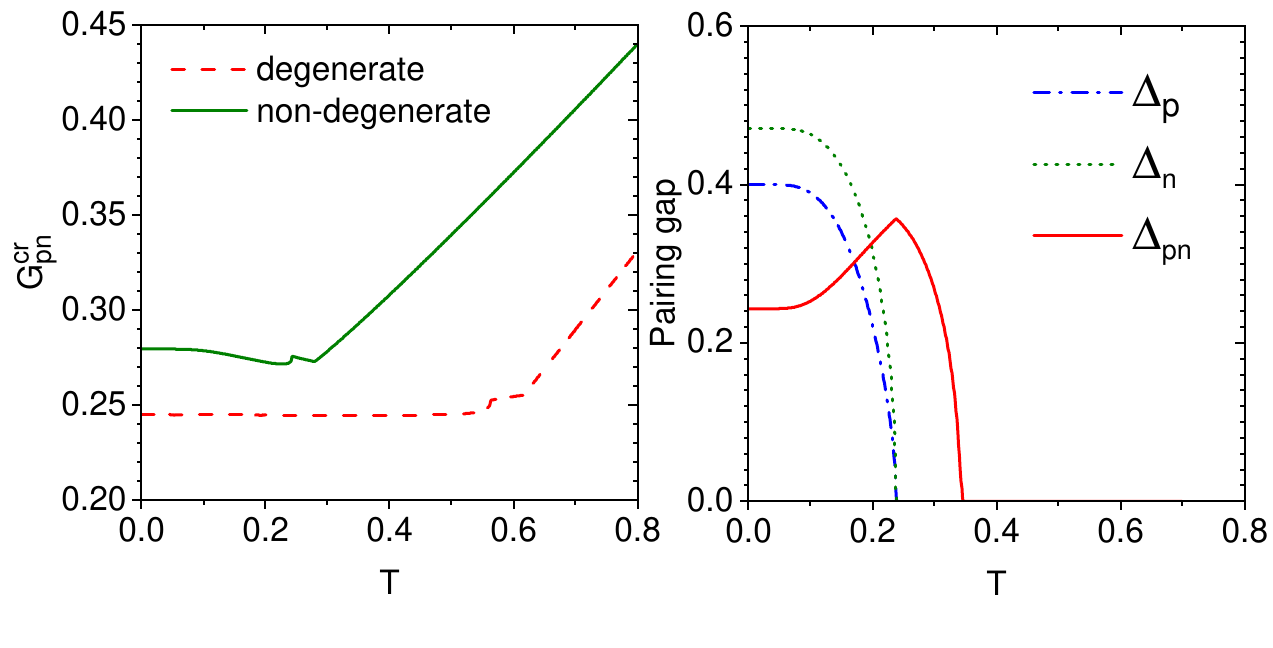}
	\caption{Left panel: Critical $pn$ pairing strength $G_{pn}^{\rm cr}$ as a function of temperature $T$ for the completely degenerate model of Ref.~\cite{PRC76_Fellah} and for the same model with identical parameters except for the introduction of non-degenerate single-particle levels, $\epsilon_{k+1} - \epsilon_k = 0.6$. Right panel: Temperature dependence of the proton, neutron and $pn$ pairing gaps in the non-degenerate model.}
	\label{fig:Gcr_degenerate}
\end{figure}

As mentioned in the Introduction, path-integral calculations~\cite{PRC76_Fellah,IJMPhE27_Mokhtari,CPC49_Fellah}  found no enhancement of $pn$ pairing at finite temperature. As shown in Fig.~2 of Ref.~\cite{PRC76_Fellah}, when the $pn$ pairing gap appears, it exhibits the conventional FT-BCS behavior, i.e., it decreases monotonically with increasing temperature without any reentrant enhancement. To explain this discrepancy with our results, we note that in Refs.~\cite{PRC76_Fellah,IJMPhE27_Mokhtari} the gap equations  identical to Eqs.~\eqref{gap2} were derived and solved for a schematic one-level model with $pp$, $nn$, and $pn$ pairing. We stress that, in contrast to the present study, the authors of Refs.~\cite{PRC76_Fellah,IJMPhE27_Mokhtari} considered a system with completely degenerate single-particle levels, i.e., $\epsilon_k = 0$ for all $k$.

In the left panel of Fig.~\ref{fig:Gcr_degenerate}, we show the temperature dependence of the critical strength $G_{pn}^{\mathrm{cr}}$ for the model parameters used in Refs.~\cite{PRC76_Fellah,IJMPhE27_Mokhtari} ($\Omega = 11$, $N_n = 6$, $N_p = 4$, and $G_p = G_n = 0.242$). As evident from the plot, for the model with completely degenerate single-particle levels, $G_{pn}^{\mathrm{cr}}$ does not exhibit any non-monotonic behavior. However, upon introducing a finite level spacing $\epsilon_{k+1} - \epsilon_k = 0.6$, $G_{pn}^{\mathrm{cr}}$ displays the same non-monotonic temperature dependence as discussed above (see Fig.~\ref{fig:Gpn_critical}). The observed decrease of $G_{pn}^{\mathrm{cr}}$ for the non-degenerate levels in certain temperature regimes implies that, for a fixed $G_{pn}$, the system can  enhance $pn$ pairing correlations.
This enhancement is illustrated in the right panel of Fig.~~\ref{fig:Gcr_degenerate}, where we use $G_{pn} = 1.2\,G_{pp}$. Note that the present calculations employ a definition of the pairing strength $G_{pn}$  that differs by a factor of two from that in Refs.~\cite{PRC76_Fellah,IJMPhE27_Mokhtari}, therefore, for a direct comparison, our results should be matched with the upper right panel ($G_{pn}=0.6G_{pp}$) of Fig.~2 in Ref.~\cite{PRC76_Fellah}.
 Thus, the reason why the path-integral calculations of Refs~\cite{PRC76_Fellah,IJMPhE27_Mokhtari} did not observe $pn$ pairing reentrance is that they employed a model with completely degenerate single-particle levels, which does not capture the thermal unblocking mechanism.

In Ref.~\cite{CPC49_Fellah}, path-integral studies were applied to the Richardson model and to realistic nuclei. In both cases, the authors assumed $N_p = N_n$. This proton--neutron symmetry leads to a typical FT-BCS behavior of the pairing gap: it exists at sufficiently large interaction strength $G_{pn}$ and then monotonically decreases with increasing temperature, exhibiting no reentrant behavior.

\subsection{Realistic nuclei \label{sec:results&realistic}}

To investigate finite-temperature $pn$  pairing correlations in Ge isotopes, we assume that nucleons move in a deformed  axially symmetric Woods–Saxon mean-field potential. Single-particle wave functions and energies are generated using the WSBETA code~\cite{CPhC46_Cwiok} with the universal parameter set (see Table~1 in~\cite{CPhC46_Cwiok}). The employed Woods-Saxon potential relaxes the charge symmetry assumption, yielding distinct single-particle energies for protons and neutrons ($\epsilon_{kp} \neq \epsilon_{kn}$). The particle model space for all nuclei is constructed up to $N=6\hbar\omega$. For the description of ground-state deformation, we include only the quadrupole deformation parameter $\beta_2$  taken from Ref.~\cite{ADNDT109_Moller}, assuming that the deformation is temperature-independent. The single-particle orbitals  are labeled by the Nilsson asymptotic quantum numbers $k=(N,\,n_z,\,\Lambda,\,\Omega)$. Due to the time-reversal invariance, states with $\Omega$ and $-\Omega$ are degenerate in energy. Pairing correlations are included only between nucleons in time-reversed states of the same orbital.

To determine the pairing strengths $G_{p}$, $G_{n}$, and $G_{pn}$ for a given nucleus, we follow the procedure outlined in Refs.~\cite{baran2005neutron,CPC49_Fellah,Lian2025}. Specifically, for each of the considered Ge isotopes the zero-temperature theoretical pairing gaps are fitted to the empirical gaps derived from experimental odd-even mass differences~\cite{vsimkovic2003proton}
\begin{align}
	\Delta_{p}^\text{emp} = &-\frac{1}{8} \left[ M(Z\!+\!2,N)\!-\!4 M(Z\!+\!1,N)\! +\! 6 M(Z,N) \right.\notag \\
	&- 4 M(Z-1,N) + M(Z-2,N) \left. \right]\,, \notag\\
	\Delta_{n}^\text{emp} = &-\frac{1}{8} \left[ M(Z,N\!+\!2)\! -\! 4 M(Z,N\!+\!1)\! +\! 6 M(Z,N) \right.\notag \\
	&- 4 M(Z,N-1) + M(Z,N-2) \left. \right]\,,\notag \\
	\Delta_{pn}^\text{emp} = &\frac{1}{4} \left\{2 \left[M(Z,N\!+\!1)\! +\! M(Z,N\!-\!1)\! +\! M(Z\!-\!1,N)\right.\right. \notag \\
	&\left. +M(Z+1,N) \right] - \left[ M(Z+1,N+1) \right.\notag \\
    & + M(Z-1,N+1)+M(Z+1,N-1) \nonumber \\
    & + M(Z-1,N-1) \left. \right] - 4 M(Z,N) \left. \right\}\,,
\end{align}
where $M(Z,N)$ are the nuclear masses~\cite{huang2021ame}. We note that since $pn$ pairing correlations influence the like-nucleon  pairing gaps and vice versa, all three pairing strengths must be determined self-consistently. The resulting values of  $G_{p}$, $G_{n}$, and $G_{pn}$, together with the empirical pairing gaps and ground-state quadrupole deformations $\beta_2$, are summarized in Table~\ref{tab1}. The table shows that the $pn$ pairing gap $\Delta_{pn}$ at $T=0$ is considerably smaller than the like-nucleon pairing gaps $\Delta_p$ and $\Delta_n$. Moreover, $\Delta_{pn}$ decreases with increasing neutron excess, in agreement with the trend reported in Refs.~\cite{civitarese1997proton,vsimkovic2003proton}.

\begin{table}[t]
	\caption{Quadrupole deformation parameters, empirical pairing gaps, and pairing strengths for selected even-even Ge isotopes.}
	\centering
	\setlength{\tabcolsep}{4.0 pt}
	\begin{tabular}{lccccccc}
		\toprule
		\midrule[0.8pt]
		 & $\beta_2$ & $\Delta_p^{\text{emp}}$ & $\Delta_n^{\text{emp}}$ & $\Delta_{pn}^{\text{emp}}$ & $G_{p}$ & $G_{n}$ & $G_{pn}$ \\
		\midrule
		$^{66}\text{Ge}$ &  0.208 & 1.607 &  1.791 & 0.778 & 0.245 & 0.242 & 0.248 \\
		$^{68}\text{Ge}$ & $-$0.276 & 1.592 & 1.868 & 0.606 & 0.241 & 0.235 & 0.254 \\
		$^{70}\text{Ge}$ & $-$0.218 & 1.545 & 1.863 & 0.588 & 0.237 & 0.222 & 0.257 \\
		$^{72}\text{Ge}$ & $-$0.217 & 1.616 & 1.835 & 0.585 & 0.239 & 0.215 & 0.269 \\
		\bottomrule
		\midrule[0.8pt]
    \label{tab1}
	\end{tabular}
\end{table}

\begin{figure}[h]
	\centering
	\includegraphics[width=0.45\textwidth]{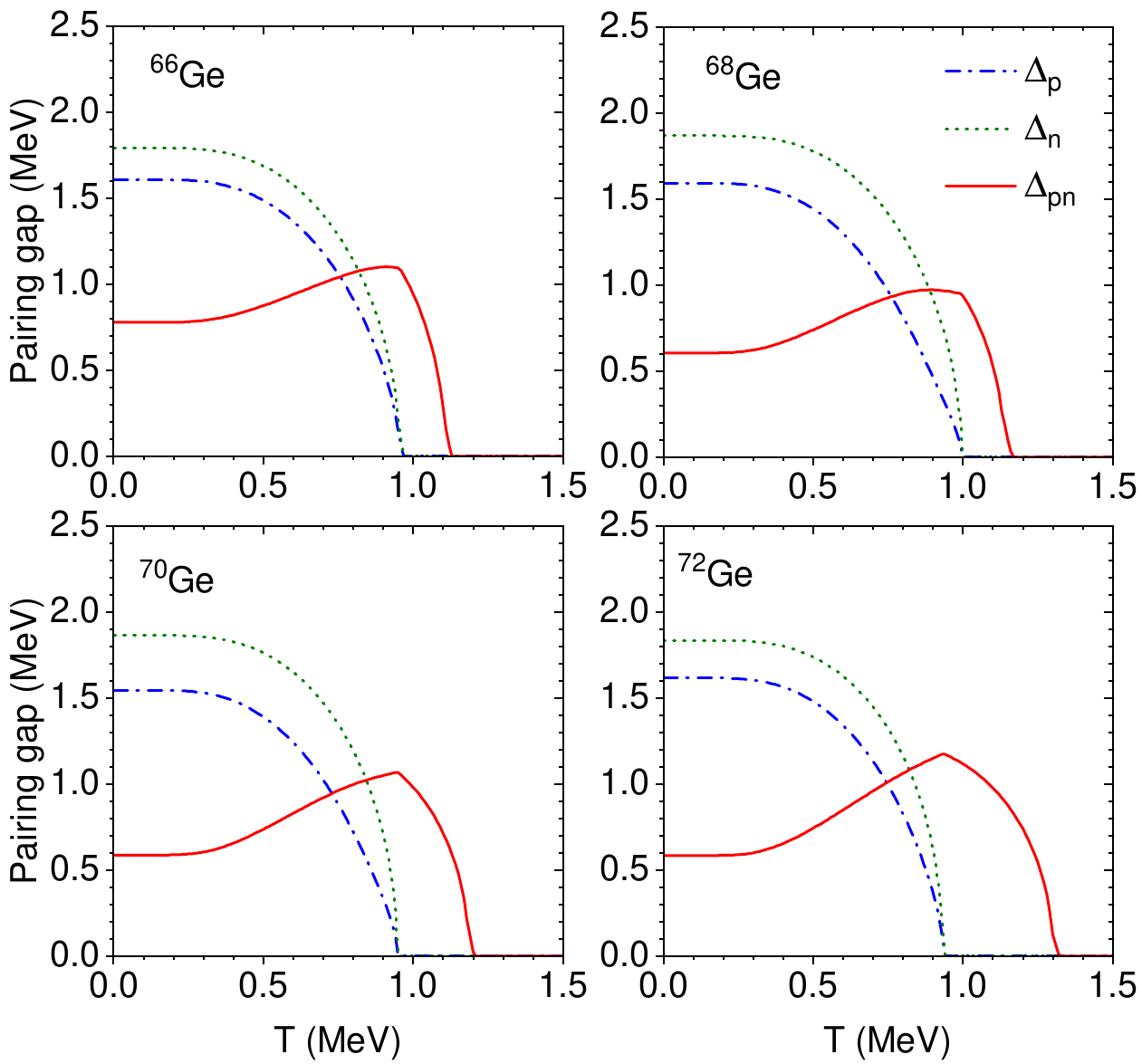}
	\caption{Temperature dependence of the proton, neutron, and proton-neutron  pairing gaps in even-even $^{66-72}$Ge isotopes.}
	\label{Ge}
\end{figure}

The temperature dependence of pairing gaps in $^{66-72}$Ge is displayed in Fig.~\ref{Ge}. One can observe that in the presence of $pn$ pairing correlations both $\Delta_p$ and $\Delta_n$ exhibit the characteristic behavior of conventional finite-temperature BCS theory: they remain nearly constant at low temperatures ($T \lesssim 0.4$\,MeV), then decrease rapidly and vanish at a critical temperature $T_\text{cr1}$. On the other hand, our calculations reveal that the $pn$ pairing reentrance phenomenon occurs in all Ge isotopes under consideration, analogous to the schematic case discussed above. While the like-nucleon pairing gaps decrease monotonically with temperature, $\Delta_{pn}$ exhibits  reentrant behavior: after  initial plateau, it rises to a maximum at $T_{\text{cr1}}$. Upon further heating, the $pn$ pairing is suppressed and $\Delta_{pn}$ vanishes at the second critical temperature $T_{\text{cr2}} > T_{\text{cr1}}$. Figure~\ref{Ge} also reveals that among the Ge isotopes considered, those with larger neutron excess, corresponding to a greater number of single-particle levels blocked for $pn$ pairing at $T=0$, exhibit a more pronounced reentrance effect: their $\Delta_{pn}$ persists to higher temperatures before being completely quenched.

\begin{figure*}[tbh!]
	\centering
	\includegraphics[width=0.8\textwidth]{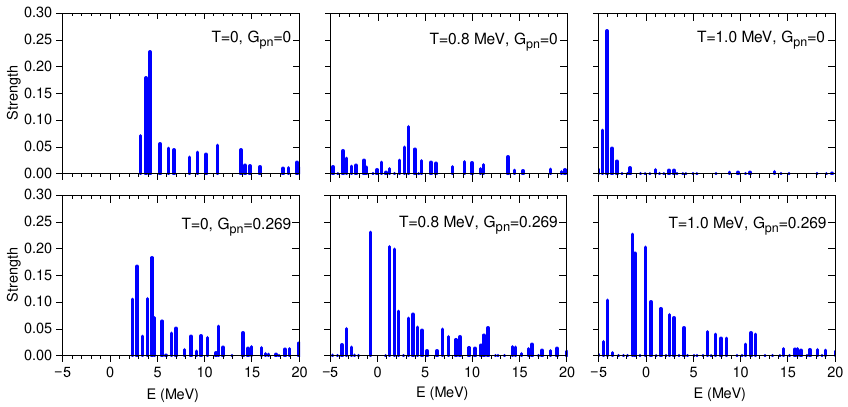}
	\caption{Strength distribution of Fermi $p\to n$ transitions in $^{72}$Ge at three temperature values. The upper row corresponds to the absence of $pn$ pairing correlations, while the lower row shows the distributions with $pn$ pairing included. The strength distributions are obtained by summing the transition strengths within 0.5\,MeV energy bins.}
	\label{72Ge_PtoN}
\end{figure*}

\begin{figure*}[tbh!]
	\centering
	\includegraphics[width=0.8\textwidth]{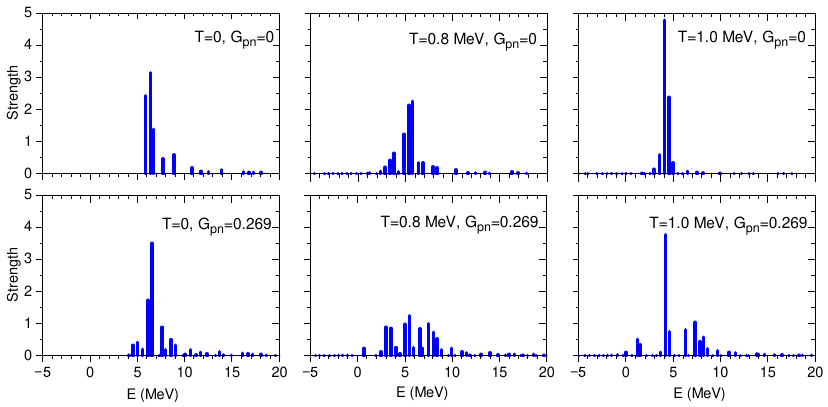}
	\caption{Same as Fig.~\ref{72Ge_PtoN} but for Fermi $n\to p$ transitions in $^{72}$Ge.}
	\label{72Ge_NtoP}
\end{figure*}

We now examine the impact of temperature-induced reentrance of $pn$  pairing correlations on the strength distribution of charge-exchange transitions in hot nuclei. Thermal modifications of the nuclear charge-exchange response function play a crucial role in determining the rates and cross sections of various weak-interaction processes such as electron/positron capture, $\beta^\pm$-decay, and (anti)neutrino absorption that occur in explosive astrophysical environments where nuclei exist at high temperatures (see Ref.~\cite{PhPN53_Dzhioev1} and references therein). In the present work, our aim is not to provide quantitatively precise strength functions but rather to perform a qualitative study of the role of  $pn$ pairing correlations. To this end, we neglect the residual interaction between thermal quasiparticles and restrict our analysis to Fermi-type transitions in nuclei with isovector $pp$, $nn$, and $pn$ pairing.

For the $n\to p$ ($p\to n$) transition, the Fermi operator reads
\begin{equation}
  F_\mp=\sum_{k,k'}\langle k p (n)|t_\mp| k' n(p)\rangle a^\dag_{k p(n)} a_{k' n(p)},
\end{equation}
where $t_-$ ($t_+$) is the isospin lowering (rising) operator.
In what follows, we adopt the approximation $\langle k p(n)|t_\mp| k' n(p)\rangle =\delta_{kk'}$. Within the independent thermal quasiparticle approximation, $F_\mp$ induces transitions from the thermal vacuum to two-quasiparticle states. To compute the corresponding transition probabilities (strengths), we expand the Fermi operators $F_\mp$ in terms of thermal quasiparticle creation operators. For the $F_-$ operator, this yields
\begin{multline}\label{F_NtoP}
   F_-\!=\! - \sum_{k>0}\sum_{\nu,\nu'}\big\{ \beta^\dag_{k\nu}\beta^\dag_{\bar k\nu'}(u_{k\nu p}v_{k\nu' n}\! +\! u_{k\nu' p}v_{k\nu n})x_{k\nu}x_{k\nu'}
              \\
 + \widetilde \beta^\dag_{k\nu}\widetilde\beta^\dag_{\bar k\nu'}(v_{k\nu p}u_{k\nu' n} + v_{k\nu' p}u_{k\nu n})y_{k\nu}y_{k\nu'}
   \\
  +\beta^\dag_{k\nu}\widetilde\beta^\dag_{k\nu'}(u_{k\nu p}u_{k\nu' n} - v_{k\nu' p}v_{k\nu n})x_{k\nu}iy_{k\nu'}
     \\
   +  \widetilde \beta^\dag_{\bar k\nu}\beta^\dag_{\bar k\nu'}(v_{k\nu p}v_{k\nu' n} - u_{k\nu' p}u_{k\nu n})iy_{k\nu}x_{k\nu'}\big\} +\ldots.
\end{multline}
Here we  omit all terms containing thermal quasiparticle annihilation operators, as they do not contribute to the strength function within the present approximation. The expression for the $F_+$ operator is obtained by interchanging the isospin indices $n$ and $p$ in the above formula.

From the expansion~\eqref{F_NtoP} it follows that the transition strength $S_i^{(\mp)}=|\langle i|F_\mp| 0(T)\rangle|^2$ from the thermal vacuum to a particular two-quasiparticle state $\ket{i}$ is given by the squared modulus of the corresponding coefficient in the expansion. For instance, the $n\to p$ transition strength to the state $\ket{i} = |\beta_{k\nu}\beta_{\bar k\nu'}\rangle$ is
\begin{equation}
S_i^{(-)} = (u_{k\nu p}v_{k\nu' n} + u_{k\nu' p}v_{k\nu n})^2x^2_{k\nu}x^2_{k\nu'}.
\end{equation}
The transition strengths $S_i^{(\pm)}$ exhibit explicit temperature dependence. Notably, transitions to states containing tilde thermal quasiparticles become possible only at finite temperature, when $y_{k\nu}\ne 0$. These correspond to thermally unblocked transitions from nuclear excited states that are forbidden at $T=0$. As discussed in Appendix~\ref{sect:superoperators}, transition strengths to tilde-conjugated states satisfy the detailed balance condition~\eqref{DetBal}, ensuring thermodynamic consistency of the strength distribution at finite temperature. The total transition strength satisfies the Ikeda sum rule $\sum_i(S^{(-)}_i - S^{(+)}_i) = N_n-N_p$.

Knowing the individual transition strengths, the finite-temperature strength function for charge-exchange transitions  can be expressed as a function of transition energy $E$
\begin{equation}
  S^{(\mp)}(E,T) = \sum_i S_i^{(\mp)} \,\delta(E-\omega_i \pm \Lambda_{np}).
\end{equation}
Here $\Lambda_{np} = \lambda_n - \lambda_p  + 1.293$\,MeV accounts for both the difference in neutron and proton chemical potentials and the neutron–proton mass splitting. The excitation energy $\omega_i$ of a two-quasiparticle state $\ket{i}$ is determined by the energies of the constituent thermal quasiparticles.  For instance, for the state $\ket{i}=|\beta_{k\nu}\beta_{\bar k\nu'}\rangle$ one has $\omega_i = E_{k\nu}+E_{k\nu'}$, while for states containing tilde quasiparticles, e.g. $\ket{i}=|\widetilde\beta_{k\nu}\widetilde\beta_{\bar k\nu'}\rangle$, the energy reads $\omega_i = -(E_{k\nu}+E_{k\nu'})$. States of mixed type such as $|\beta_{k\nu}\widetilde\beta_{k\nu'}\rangle$ have excitation energies $\omega_i = E_{k\nu}-E_{k\nu'}$. These relations follow directly from the diagonal form of the thermal Hamiltonian~\eqref{H_thermal_diagonal}. At finite temperature, the transition energy $E$
can take negative values corresponding to thermally unblocked de-excitation processes that are forbidden at zero temperature.

In Figs.~\ref{72Ge_PtoN} and \ref{72Ge_NtoP} we present the strength functions for Fermi $p\to n$ and $n\to p$ transitions in $^{72}$Ge calculated without ($G_{pn}=0$) and with ($G_{pn}=0.269$\,MeV) $pn$ pairing. The distributions are shown at three characteristic temperatures: at $T=0$ the $pn$ pairing gap is small; at $T=0.8$\,MeV the $pn$ gap reaches approximately half the sum of the like-nucleon gaps ($\Delta_{pn}\approx 0.5\,(\Delta_p+\Delta_n)$), indicating that $pn$ pairing becomes comparable in strength to like-nucleon pairing; and at $T=1.0$\,MeV the like-nucelon pairing vanishes while $\Delta_{pn}$ approaches its maximum value. Note that the temperature regime around $T\approx 1$\,MeV is typically encountered in the collapsing core of a massive star during the late silicon-burning and early collapse stages, as well as in the envelope of accreting neutron stars during Type I X-ray bursts, which is the astrophysical site where the rapid proton-capture (rp) process operates.

As it clearly seen in Fig.~\ref{72Ge_PtoN}, at $T=0$ the  $p\to n$ strength distributions obtained with and without $pn$ pairing are qualitatively similar:
they are concentrated in the energy interval $2.5-5$\,MeV, but the inclusion of $pn$ pairing slightly redistributes strength toward lower energies.
At $T=0.8$\,MeV, the distribution with $pn$ pairing shows enhanced low‑ and negative-energy strength compared to the $G_{pn}=0$ case. The fragmentation pattern changes completely at $T=1.0$\,MeV, reflecting the increasing influence of $pn$ correlations on the thermal quasiparticle spectra and structure. The strength function  without $pn$ pairing exhibits a pronounced concentration at negative energies in stark contrast to the broad distribution obtained with $pn$ pairing. This demonstrates that  reentrance of $pn$ correlations can qualitatively reshape the $p\to n$ charge‑exchange response even when conventional pairing is fully quenched. Taking into account that this response governs not only supernova electron capture but also $\beta^+$-decay of nuclei formed in the rapid proton capture process, the reentrance effect could modify the neutronization rate of stellar matter and influence the reaction flow along the $rp$-process path.

The $n\to p$ strength functions in Fig.~\ref{72Ge_NtoP} display an analogous temperature dependence but with notable asymmetries relative to $p\to n$ strength, reflecting the sensitivity of the response to the isospin structure of the thermal quasiparticle wave functions. First of all, the overall magnitude is larger due to the neutron excess in $^{72}$Ge. Second, the energy centroid of the $n\to p$ distribution is shifted toward higher excitation energies, a consequence of the negative value of $\Lambda_{np}$ in $^{72}$Ge. For this reason,   the bulk of the thermally unblocked  $n\to p$ strength  is concentrated at low positive energies, while the strength at negative energies (corresponding to de-excitation processes) remains relatively small.
However, like in $p\to n $ case, at  $T=1.0$\,MeV  we observe a dramatic effect of $pn$ pairing reentrance:
the $n\to p$ strength is concentrated in a strong peak at $E\approx 4$\,MeV when $pn$ pairing is neglected, whereas the $G_{pn}\ne 0$ calculation yields low- and high-energy tails.

\begin{figure}[tbh!]
	\centering
	\includegraphics[width=\columnwidth]{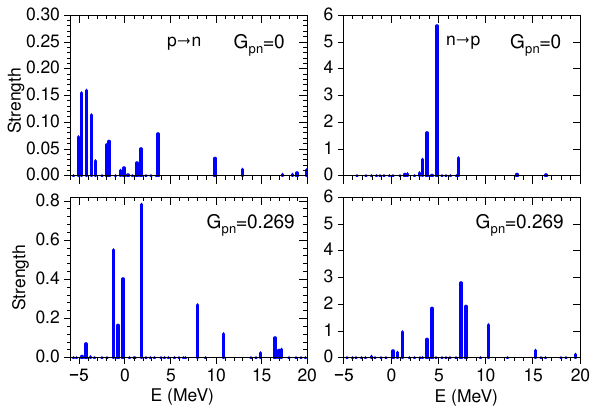}
	\caption{Same as Figs.~\ref{72Ge_PtoN} and \ref{72Ge_NtoP}, but for Fermi transitions in spherical $^{72}$Ge at $T=1.0$\, MeV.}	\label{72Ge_spherical}
\end{figure}

The present qualitative study underscores the need for further investigations into the role of $pn$ pairing in determining stellar weak-interaction observables under extreme astrophysical conditions.  A crucial next step in this direction is the inclusion of temperature-dependent deformation, which would require fully self-consistent calculations of the mean-field and pairing correlations at finite temperature.  To address the uncertainty introduced by the assumption of a temperature-independent shape, and to demonstrate that a possible shape transition around the temperatures where $pn$-pairing reentrance occurs does not alter our main qualitative conclusions, we performed an additional test calculation. In Fig.~\ref{72Ge_spherical}, we show the Fermi strength distributions in $^{72}$Ge at $T=1.0$\,MeV computed using a spherically symmetric mean-field ($\beta_2=0.0$), while keeping all other model parameters identical to the deformed case.

As follows from our calculations, even in the spherical mean-field at $T=1.0$\,MeV, the like-nucleon $pp$ and $nn$ pairing correlations are completely quenched, whereas the $pn$ pairing gap survives and reaches a value comparable to the deformed case ($\Delta_{pn} \approx 1.3$\,MeV). More importantly, the inclusion of $pn$ pairing produces the same profound qualitative effect on the Fermi strength functions as observed in the deformed nucleus. Specifically, for the $n \rightarrow p$ transitions (right panels of Fig.~\ref{72Ge_spherical}), the sharp, highly concentrated peak located at $E \approx 5$\,MeV in the absence of $pn$ pairing is completely fragmented into a broad distribution with distinct peaks at both lower and higher energies. Similarly, the $p \rightarrow n$ strength distribution (left panels) is noticeably reshaped and redistributed across the energy spectrum.

These results suggest that the significant impact of $pn$ pairing reentrance on the charge-exchange response is likely a general feature, rather than an artifact of a specific deformation assumption.  Consequently, while quantitative astrophysical rates will eventually require the inclusion of temperature-dependent deformation (and residual interactions), the qualitative modifications to the strength functions discussed above are expected to persist regardless of shape evolution.

\section{CONCLUSIONS \label{sec:conclusions}}

In the present work, we have developed a generalized finite-temperature proton-neutron BCS (FT-pnBCS) framework for both isovector and isoscalar  pairing channels using the superoperator formalism. Numerical calculations for a schematic doubly degenerate equidistant multilevel pairing model as well as for realistic even-even Ge isotopes demonstrate the emergence of $pn$ pairing reentrance in  $N>Z$ even-even nuclei with preexisting like-nucleon pairing. This phenomenon is driven by thermal excitations, which reduce Pauli blocking through the partial unblocking of single-particle levels that otherwise suppress $pn$ correlations at low temperatures.

The reentrance phenomenon we observe occurs in the temperature regime $T\approx 1$\,MeV, where the mean-field FT-BCS approximation predicts a sharp phase transition and both thermal fluctuations and particle-number projection effects may become non-negligible.  However, as discussed in the Introduction, the temperature-induced $pn$ pairing reentrance  due to unblocking of single-particle levels near the Fermi surface has been demonstrated in odd--odd~\cite{kaneko2005proton} and rotating~\cite{sheikh2005reappearance} $N=Z$ nuclear systems using particle number conserving approaches that avoid sharp phase transitions entirely. This provides strong evidence that thermally-induced $pn$ pairing reentrance is a robust qualitative feature that persists beyond the mean-field approximation, regardless of whether the low-temperature blocking arises from an odd/unpaired nucleon or from the neutron excess. We acknowledge, however, that quantitative aspects, such as the sharpness of the $pn$ pairing enhancement and the precise boundaries of the reentrant temperature window, may be modified by a particle-number conserving shell-model Monte-Carlo treatment~\cite{KooninPhysRep278} or by projection techniques (e.g., Lipkin-Nogami~\cite{Kosov1996,DangPRC47}). This constitutes an important direction for future investigations.

Another limitation of the FT-pnBCS framework is that it does not allow the simultaneous presence of isoscalar ($t=0$) and isovector ($t=1$) pairing fields when their coupling strengths are different. The possibility of coexistence of isovector and isoscalar pairing correlations in realistic nuclei remains a subject of ongoing theoretical and experimental debate~\cite{PPNP78_Frauendorf}. Consequently, the specific reentrance behavior reported here should be regarded as a model-dependent prediction. In this respect, an extension of the present FT-pnBCS approach incorporating particle-number projection, similar to that developed at zero temperature~\cite{PLB393_Satula}, would provide a natural framework for investigating the coexistence and competition of isovector and isoscalar pairing correlations at finite temperature. Nevertheless, the reentrance mechanism discussed here is driven primarily by the thermal unblocking of single-particle levels and this mechanism is not specific to the isospin character of the pairing channel.  This suggests that the thermal-unblocking mechanism should remain operative even within more general theoretical frameworks, although the quantitative characteristics of the reentrance phenomenon may depend on the treatment of competing pairing channels.

The qualitative analysis of Fermi charge-exchange strength distributions in hot $^{72}$Ge performed within independent thermal quasiparticle approximation motivates further systematic studies to clarify how $pn$ pairing correlations modify stellar weak-interaction observables in high-temperature astrophysical environments. Besides the inclusion of temperature-dependent deformation, this will
also require taking into account residual interactions, as was done for spherical
nuclei at finite temperature within the thermal quasiparticle RPA~\cite{PhPN53_Dzhioev1,*PhPN53_Dzhioev2,*PhPN53_Dzhioev3}.

The impact of $pn$ pairing reentrance on key nuclear observables such as the level density, $\gamma$-ray strength function, and $\beta$-decay rates of thermally excited nuclei will be addressed in future works. Furthermore, the present formalism can naturally be extended to hot, rapidly rotating nuclei, thereby incorporating additional collective degrees of freedom.

\begin{acknowledgments}
  This work was funded by the National Foundation for Science and Technology Development (NAFOSTED) of Vietnam (Grant No. 103.04-2025.22).
\end{acknowledgments}

\appendix

\section{Overview of the superoperator formalism}\label{sect:superoperators}

One of the principal technical complications in statistical quantum mechanics arises from the fact that systems are described by the  density operator
$\rho$ rather than by a state vector. The superoperator formalism offers a mathematical framework that restores a state-vector description for statistical systems.  In this Appendix, we briefly discuss details of the formalism that are relevant to our paper, while the comprehensive description is given in~\cite{PhPN53_Dzhioev1}.

Consider a quantum system described by a Hamiltonian \( H \) acting on the Hilbert space \( \mathfrak{H} \).  The mixed state of the system is  characterized by the density operator $\rho$, which can be expanded in an orthonormal basis  \( \{ |n\rangle \} \) of \( \mathfrak{H} \) as
\begin{equation}\label{def_rho}
\rho = \sum_{nm}\rho_{nm}|n\rangle\langle m|,
\end{equation}
where \( \rho_{nm}=\langle n|\rho|m\rangle \) are the matrix elements.  The time evolution of the density operator is governed by the Liouville–von Neumann equation:
\begin{equation}\label{LvN}
\frac{d\rho}{dt} = -i[H, \rho].
\end{equation}

At thermal equilibrium and finite temperature $T$, the mixed state is  characterized by the Gibbs density operator
\begin{equation}\label{def_rho_eq}
{\rho}(T) = \frac{e^{-H/T}}{Z(T)},
\end{equation}
where the partition function \( Z(T) = \operatorname{Tr}(e^{-H/T})\) ensures the normalization of the density operator \( \operatorname{Tr}{\rho} = 1 \). The time-independent equilibrium density operator~\eqref{def_rho_eq}  is a stationary solution of the Liouville-von Neumann equation and it satisfies the Kubo-Martin-Schwinger (KMS) boundary condition (see, e.g., Sect.~2.2.e in Ref.~\cite{Emch1974}). Specifically, for any two
Heisenberg-picture operators $A(t)$ and $B(t')$, the equilibrium correlation
functions at inverse temperature $\beta = 1/T$ obey the relation
\begin{equation}\label{KMS}
    \bbrakett{A(t)\,B(t')} = \bbrakett{B(t')\,A(t + i\beta)},
\end{equation}
where the double-bracket notation denotes the statistical average with respect to $\rho(T)$:
\begin{equation}\label{thermal_avg}
    \bbrakett{X} \equiv \operatorname{Tr}\bigl(\rho(T)\,X\bigr).
\end{equation}

The cornerstone of the superoperator formalism is the isomorphism $\mathfrak{L}(\mathfrak{H})\cong \mathfrak{H}\otimes \mathfrak{H}^*$ between the Liouville space $\mathfrak{L}$ of linear operators on the Hilbert space $\mathfrak{H}$ and vectors in the extended Hilbert space   $\mathfrak{H}\otimes \mathfrak{H}^*$~\footnote{Such an isomorphism is often called  the Choi-Jamiolkowski isomorphism between states and operators~\cite{JPhA57_Frembs}.}. Here $\mathfrak{H}^*$ denotes the dual space to $\mathfrak{H}$, i.e. it is the space of bras instead of the space of kets.   Thus, given an orthonormal basis \( \{ |n\rangle \} \) in the Hilbert space \( \mathfrak{H} \), the operators \( \kett{n m} \equiv |n\rangle\langle m| \) form a complete orthonormal basis in $\mathfrak{L}(\mathfrak{H})$.  This correspondence allows us to represent linear operators on \( \mathfrak{H} \) as vectors (pure states) in the  associated  Liouville space.  In particular, the density operator \eqref{def_rho} is mapped to the vector
 \begin{equation}\label{def_rho_vec}
     \kett{\rho} = \sum_{m,n} \rho_{nm}\, \kett{nm}.
 \end{equation}
 Superoperators are linear transformations acting on the Liouville space. The central object in this formalism is the Liouvillian superoperator
\( \mathcal{L} \), defined by its action on the vectorized density operator:
\(\mathcal L\kett{\rho}=\kett{[H,\rho]}\). Within the superoperator formalism, the Liouville-von Neumann equation can be
cast into a Schr\"odinger-like form:
\begin{equation}\label{LvN_super}
    \frac{d}{dt}\,\kett{\rho(t)} = -i\,\mathcal{L}\,\kett{\rho(t)}.
\end{equation}
Consequently, the thermal equilibrium state satisfies the stationary condition
\begin{equation}\label{stationary}
    \mathcal{L}\,\kett{\rho(T)} = 0.
\end{equation}

 To construct an explicit representation of the Liouvillian superoperator \( \mathcal{L} \), one must first specify a basis in the Hilbert space \( \mathfrak{H} \). For the nuclear many-body problem, a natural choice is the set of particle-number eigenstates \( \{ |n\rangle \} \) defined by   \(a^\dagger_k a_k \, |n\rangle = n_k \, |n\rangle \),
 where $a_k^\dagger$ and $a_k$ denote, respectively, the particle creation and annihilation operators, and \( n_k \in \{0,1\} \) denotes the occupation number of the single-particle state \( k \).
 It was shown in Ref.~\cite{PhPN53_Dzhioev1} that once the nuclear Hamiltonian \( H(a^\dagger, a) \) is given in second-quantized form the Liouvillian superoperator can be expressed as
 \begin{equation}
      \mathcal{L} = H(\vec{a}^\dagger, \vec{a}) - H(\cev{a}^\dagger, \cev{a}),
 \end{equation}
 where \( \vec{a}^\dagger, \vec{a} \) and \( \cev{a}^\dagger, \cev{a} \) denote the left- and right-action creation and annihilation superoperators, respectively. Their action on the vectorized basis states \( \kett{nm}\) is defined by
 \begin{align}
  &  \vec a^\dag_k\kett{nm}\equiv a^\dag_k\ket{n}\!\bra{m},~~ \cev a^\dag_k\kett{nm}\equiv\alpha_{mn} \ket{n}\!\bra{m} a_k
    \notag\\
  & \vec a_k\kett{nm}\equiv a_k\ket{n}\!\bra{m},~~ \cev a_k\kett{nm}\equiv\alpha_{mn}\ket{n}\!\bra{m} a^\dag_k,
 \end{align}
 where  the phase factor \( \alpha_{nm} = i\,(-1)^{N_n + N_m} \), with \( N_n = \sum_k n_k \) being the total particle number in the state \( |n\rangle \), ensures the correct fermionic structure of the algebra generated by the creation and annihilation superoperators in the Liouville space.  Thus, within the superoperator framework, the original set of fermionic creation and annihilation operators is effectively doubled, while their
canonical anticommutation relations are preserved. This extension of the second quantization to the Liouville space is sometimes referred to in
the literature as the third quantization~\cite{NJPh10_Prosen}.

In what follows, for convenience, we adopt the notation and terminology of Thermo Field Dynamics (TFD)~\cite{takahashi1996thermo}.\footnote{The connection between TFD and the superoperator formalism was established by Schmutz~\cite{schmutz1978real} and is further discussed in Ref.~\cite{PhPN53_Dzhioev1}.}
In the TFD notation, the left-action superoperators correspond to the physical  operators acting on the original (physical) Hilbert space, whereas the right-action superoperators are associated with their counterparts acting on the fictitious (tilde) copy of the Hilbert space. In other words, within the TFD framework, every physical operator \( X = X(a^\dagger, a) \) possesses a tilde counterpart  \(\widetilde{X} = X^*(\widetilde{a}^\dagger, \widetilde{a})\)
 obtained from \( X \) by the tilde conjugation operation: each nucleon creation (annihilation) operator is replaced by its tilde partner (\( a \mapsto \widetilde{a} \), \( a^\dagger \mapsto \widetilde{a}^\dagger \)), and all complex coefficients are complex-conjugated (\( c \mapsto c^* \)).

In TFD, the equilibrium density matrix \( {\rho}(T) \) is encoded as a temperature-dependent vector, referred to as the thermal vacuum \( |0(T)\rangle \) and the statistical average of any observable \( X \) takes the form of a vacuum expectation value:
\begin{equation}
    \bbrakett{X} = \langle 0(T) | X | 0(T) \rangle.
\end{equation}
The Liouvillian superoperator $\mathcal{L}$ is identified with the thermal Hamiltonian \( \mathcal{H} \):
\begin{equation}
     \mathcal{H} = H - \widetilde{H},
\end{equation}
and the stationarity condition~\eqref{stationary} for thermal equilibrium is equivalently expressed as
\begin{equation}
    \mathcal{H} \, |0(T)\rangle = 0.
\end{equation}
The KMS condition~\eqref{KMS} finds its counterpart in TFD notation as the thermal-state condition (TSC):
\begin{equation}\label{TSC}
    X|0(T)\rangle = \sigma_X e^{\mathcal{H}/(2T)} \widetilde{X}^\dag \ket{0(T)},
\end{equation}
where the phase factor \( \sigma_X \) equals \( +1 \) for bosonic-like operators and \( -i \) for fermionic-like operators.\footnote{
An operator \( X(a^\dagger, a) \) is termed bosonic-like (fermionic-like) if every term in its second-quantized expansion contains an even (odd) number of fermionic creation and annihilation operators.
 It should be noted that in the original formulation of TFD~\cite{takahashi1996thermo} the phase factor \( \sigma_X \)
 depends in a nontrivial way on the structure of the operator \( X \) reflecting its particle-number parity and statistical nature.
 However, as shown in Ref.~\cite{PhPN53_Dzhioev1}, this dependence can be significantly simplified by suitably redefining fermionic creation and annihilation superoperators.
 Moreover, this redefinition ensures that tilde conjugation is an involution, i.e., \( \widetilde{\widetilde{X}} = X \). }

 Thus, within the superoperator formalism, the construction of the thermal vacuum and the evaluation of statistical averages is reduced to finding the zero-eigenvalue
eigenstate of the thermal Hamiltonian \( \mathcal{H} \) that satisfies the thermal-state condition~\eqref{TSC}. Eigenstates of \( \mathcal{H} \) with
nonzero eigenvalues describe non-equilibrium states of the system at finite temperature induced by external perturbations. This interpretation follows from
the fact that \( \mathcal{H} \) generates time evolution in the Liouville space. The presence of two-body interactions in \( \mathcal{H} \) generally precludes
exact diagonalization; however, an approximate diagonal form can be achieved within a suitable many-body scheme:
\begin{equation}\label{H_diag}
    \mathcal{H} \approx \sum_i \omega_i \bigl( \mathcal{O}_i^\dagger \mathcal{O}_i
    - \widetilde{\mathcal{O}}_i^\dagger \widetilde{\mathcal{O}}_i \bigr).
\end{equation}
Note that at finite temperature both excitation and de-excitation processes become physically accessible, corresponding respectively to the positive- and
negative-energy eigenmodes of \( \mathcal{H} \).

Once the thermal Hamiltonian is diagonalized within a given approximation, it is natural, by a direct analogy with the zero-temperature case, to define the
corresponding (approximate) thermal vacuum \( |0(T)\rangle \) as a vacuum state for the annihilation superoperators:
\begin{equation}\label{annihilation}
    \mathcal{O}_i \, |0(T)\rangle = 0, \qquad
    \widetilde{\mathcal{O}}_i \, |0(T)\rangle = 0.
\end{equation}
The thermal-state condition~\eqref{TSC} then constrains not only the equilibrium state but also the structure of non-equilibrium excitations. In particular, it
enforces a detailed-balance relation between excitation and de-excitation amplitudes: for any transition operator \( \mathcal{T} \),
\begin{equation}\label{DetBal}
    \bigl| \langle \widetilde{\mathcal{O}}_i | \mathcal{T} | 0(T) \rangle \bigr|^2
    = e^{-\omega_i / T} \,
      \bigl| \langle \mathcal{O}_i | \mathcal{T}^\dagger | 0(T) \rangle \bigr|^2.
\end{equation}
In this way, the superoperator formalism offers a thermodynamically consistent framework for describing both equilibrium and non-equilibrium properties of
atomic nuclei at finite temperature within a suitable many-body approximation. Following the conceptual framework of the quasiparticle-phonon nuclear
model~\cite{Soloviev_QPM}, Ref.~\cite{PhPN53_Dzhioev1} provides a detailed description of three successive many-body schemes for diagonalizing the thermal
Hamiltonian: (i) thermal quasiparticles, (ii) thermal phonons, and (iii) the coupling between thermal quasiparticles and thermal phonons.

%\bibliographystyle{apsrev4-2}
%\bibliography{references-new.bib}
%\end{document}

%

\end{document}